\documentclass[10pt]{iopart}

\usepackage[square]{natbib}
\usepackage{amssymb}

\expandafter\let\csname equation*\endcsname\relax
\expandafter\let\csname endequation*\endcsname\relax
\usepackage{amsmath}
\usepackage{lineno}
\usepackage{gensymb}
\usepackage{graphicx}
\usepackage{color}

\newcommand{\unit}[1]{\ensuremath{\, \mathrm{#1}}}

\newcommand*\patchAmsMathEnvironmentForLineno[1]{%
  \expandafter\let\csname old#1\expandafter\endcsname\csname #1\endcsname
  \expandafter\let\csname oldend#1\expandafter\endcsname\csname end#1\endcsname
  \renewenvironment{#1}%
     {\linenomath\csname old#1\endcsname}%
     {\csname oldend#1\endcsname\endlinenomath}}%
\newcommand*\patchBothAmsMathEnvironmentsForLineno[1]{%
  \patchAmsMathEnvironmentForLineno{#1}%
  \patchAmsMathEnvironmentForLineno{#1*}}%
\AtBeginDocument{%
\patchBothAmsMathEnvironmentsForLineno{equation}%
\patchBothAmsMathEnvironmentsForLineno{align}%
\patchBothAmsMathEnvironmentsForLineno{flalign}%
\patchBothAmsMathEnvironmentsForLineno{alignat}%
\patchBothAmsMathEnvironmentsForLineno{gather}%
\patchBothAmsMathEnvironmentsForLineno{multline}%
}

\newcommand*\newblock{\hskip .11em\@plus.33em\@minus.07em}

\begin{document}

\title[]{Cumulant expansions for atmospheric flows}
\author{Farid Ait Chaalal$^1$,  Tapio Schneider$^{1,2}$, Bettina Meyer$^1$, \\ and Brad Marston$^3$}

\address{$^1$ETH Z\"urich, Zurich, Switzerland \\
         $^2$California Institute of Technology, Pasadena, California \\
         $^3$Brown University, Providence, Rhode Island}
         

\begin{abstract}

Atmospheric flows are governed by the equations of fluid dynamics. These equations are nonlinear, and consequently the hierarchy of cumulant equations is not closed. But because atmospheric flows are inhomogeneous and anisotropic, the nonlinearity may manifest itself only weakly through interactions of non-trivial mean fields with disturbances such as thermals or eddies. In such situations, truncations of the hierarchy of cumulant equations hold promise as a closure strategy.

Here we show how truncations at second order can be used to model and elucidate the dynamics of turbulent atmospheric flows. Two examples are considered. First, we study the growth of a dry convective boundary layer, which is heated from below, leading to turbulent upward energy transport and growth of the boundary layer. We demonstrate that a quasilinear truncation of the equations of motion, in which interactions of disturbances among each other are neglected but interactions with mean fields are taken into account, can capture the growth of the convective boundary layer. However, it does not capture important turbulent transport terms in the turbulence kinetic energy budget. Second, we study the evolution of two-dimensional large-scale waves, which are representative of waves seen in Earth's upper atmosphere. We demonstrate that a cumulant expansion truncated at second order (CE2) can capture the evolution of such waves and their nonlinear interaction with the mean flow in some circumstances, for example, when the wave amplitude is small enough or the planetary rotation rate is large enough. However, CE2 fails to capture the flow evolution when strongly nonlinear eddy--eddy interactions that generate small-scale filaments in surf zones around critical layers become important. Higher-order closures can capture these missing interactions.

The results point to new ways in which the dynamics of turbulent boundary layers may be represented in climate models, and they illustrate different classes of nonlinear processes that can control wave dissipation and angular momentum fluxes in the upper troposphere.

\end{abstract}

%
\vspace{2pc}
%
\submitto{\NJP}
%
%
%

\section{Introduction}\label{sec:introduction}

Atmospheric flows shape Earth's climate and are governed by the equations of fluid dynamics, the Navier-Stokes equations augmented by the Coriolis force and thermodynamic equations \citep[e.g.,][]{ooyama2001,vallis2006,pauluis2008}, and equations for the microphysical processes describing, for example, the formation and re-evaporation of cloud droplets \citep{pruppacher1998}. They span an enormous range of length scales, from the micrometers of droplet formation to the planetary scale. Temporal variations range from microseconds at the smallest scales to tens of years on the largest scales  \citep[e.g.,][]{klein2010}. Atmospheric processes are tightly coupled across all of these scales. For example, cloud droplets scatter sunlight and absorb infrared radiation, thereby affecting Earth's radiative balance globally; conversely, planetary-scale dynamics affect where and how clouds form. Current climate models cannot resolve all relevant scales. They resort to the direct simulation of dynamics on scales of tens of kilometers and larger, while representing smaller-scale processes such as turbulence in clouds and boundary layers semi-empirically \citep[e.g.,][]{beljaars92a,garratt1994,smith1997,lappen01a,soares04,Siebesma07}. However, the larger-scale dynamics of weather systems, with timescales of minutes, are simulated explicitly even when only their longer-term statistics---the climate---is ultimately of interest. 

Two scientific objectives would be beneficial to achieve. First, it would be desirable to obtain more accurate and more physically motivated models of the interactions between the larger scales that can currently be resolved in climate models and the smaller scales that cannot be resolved. Inaccuracies in how these interactions, in particular in clouds and boundary layers, are represented in climate models are the largest source of uncertainties in climate projections \citep[e.g.,][]{stephens2005,bony2006,soden2006,webb2013,stevens2013,vial2013,brient2016}. Improving the representation of such interactions would have an enormous societal benefit. Second, it would be desirable to devise ways of calculating climate statistics more directly, rather than through the direct simulation of weather systems and accumulation of their statistics, as is currently done. This may in the long run lead to faster climate simulations. In the shorter term, it may lead to insight into how climate is maintained and how it varies on timescales of seasons to millennia. 

Both objectives require the development of closure models for the hierarchy of statistical moment or cumulant equations associated with the equations of fluid dynamics. This hierarchy is in principle infinite because of the quadratic nonlinearity of the Navier-Stokes equations. Numerous ways of closing the hierarchy of moment or cumulant equations in a variety of circumstances have been proposed \citep[see, e.g.,][]{frisch1995,Pope00a,lesieur2008}. Many of them concern flows that are assumed statistically homogeneous and isotropic, as an idealized benchmark from which the development of closures for more realistic applications can proceed \citep[e.g.,][]{Orszag70a,Orszag73a}. However, closures for homogeneous and isotropic turbulence often may not be easier to obtain than closures for more realistic flows: Mean fields in homogeneous and isotropic turbulence can, without loss of generality, be taken to be zero; only higher-order statistics of the flows are of interest. Isotropic turbulence cannot equilibrate with any imposed driving and dissipation through interaction with mean flows; rather, it must equilibrate through nonlinear interactions across scales. By contrast, turbulence in the atmosphere usually interacts strongly with non-trivial mean fields, which include, for example, atmospheric jet streams or the thermal stratification of the atmosphere. Because interactions between turbulent fluctuations and non-trivial mean fields have the potential to be important, many atmospheric flows may be less strongly nonlinear than the oft-studied prototype problems of three-dimensional turbulence \citep[e.g.,][]{pedlosky1970,farrell1987,farrell1993,randel1991,schneider2006}. Moreover, already the mean fields (e.g., mean temperatures and winds) are of primary interest for understanding climate, though, of course, higher moments (e.g., temperature extremes) also remain important to understand. 

Because the nonlinearity of turbulent interactions in many atmospheric flows may be limited, truncating the hierarchy of moment or cumulant equations at a low order has potential to be successful. Here we explore the feasibility of truncations at second order---that is, neglecting third-order nonlinearities in second-order covariance equations---in two prototype problems of atmospheric flows. The first is a turbulent convective boundary layer, with scales of motion on the order of meters to a kilometer. The second is a model of large-scale turbulence in the upper atmosphere, with scales of motion on the order of hundreds to thousands of kilometers. These two problems involve disparate phenomenologies and force balances. For example, the boundary layer can be taken to be unaffected by the planetary rotation, whereas the planetary rotation and Coriolis forces are fundamental for the large-scale turbulence in the upper atmosphere. Yet the problems share that turbulent fluctuations interact strongly with a non-trivial mean state---a thermal stratification in the first case, and an atmospheric jet in the second case. Because of the strength of this interaction, truncations of moment equations at second order already achieve some success in capturing the statistics of these flows. It is essential in these truncations that nonlocal and anisotropic covariation of turbulent quantities (e.g., in waves or convective plumes) are retained. Such nonlocal truncation of the moment or cumulant equations at second order is known as second-order cumulant expansion (CE2) \citep{marston2008,marston2011,srinivasan2012,tobias2013} or stochastic structural stability Theory (S3T) \citep{farrell2003,farrell2007,bakas2014,constantinou2014b}. CE2 and S3T differ in that S3T attempts to represent missing eddy-eddy interactions, whereas CE2 sets them to zero. 

CE2 is a realizable closure in that its equations are the exact moment equations of a realizable system, the quasi-linear (QL) system that corresponds to the original equations of motion. The QL approximation retains the interaction of turbulent fluctuations with a mean flow but neglects the interactions of turbulent fluctuations among each others. CE2 and S3T were successful in explaining some aspects of zonal jet dynamics in rotating flows \citep[e.g.,][]{farrell2003,farrell2007,srinivasan2012,tobias2013,bakas2013}, without relying on eddy--eddy interactions and inverse cascades \citep{rhines1975,vallis1993}. QL approximations of large-scale atmospheric dynamics, sometimes with added damping and stochastic forcing, were partially successful in reproducing aspects of the atmospheric climate and its variability \citep[e.g.,][]{whitaker1998,zhang1999,delsole2001,ogorman2007}. At smaller scales, QL approximations also capture sheared stably stratified flows when the dynamics involve the linear excitation and absorption of internal gravity waves \citep{orr1907,lindzen1988,bakas2007}. They also reproduce aspects of thermal convection, such as the dependence of the heat flux on the Rayleigh number \citep[e.g.,][]{malkus1954, herring1963, Toomre77a, Busse78, niemela2000}.

In what follows, we derive the CE2 closure for Boussinesq flows, present fully nonlinear and QL simulations of a dry convective boundary layer using large-eddy simulations (LES), and study the evolution of a large-scale wave disturbance on a zonal jet representative of the upper troposphere. The results will demonstrate the potential and limitations of CE2 approaches.

\section{Cumulant expansion of Boussinesq flow}\label{sec:CE2}

\subsection{Boussinesq flow}

Atmospheric flows have low Mach number, so sound waves are generally unimportant for the dynamics. It is therefore common to study atmospheric dynamics with approximations to the Navier-Stokes equations that filter sound waves. The simplest such approximation, which ignores all density variations except where they affect the buoyancy of air masses, is the Boussinesq approximation \citep{Boussinesq1872}. The Boussinesq equations are obtained by expressing density $\rho(\mathbf{r},t) = \rho_0 + \delta\rho(\mathbf{r},t)$ as a sum of a constant density $\rho_0$ in a reference state and fluctuations $\delta\rho(\mathbf{r},t)$ about it, assuming pressure variations in the reference state are hydrostatically balanced, and retaining only the leading-order terms in density and pressure fluctuations in an expansion of the Navier-Stokes equations. The resulting equations are \citep[e.g.,][]{vallis2006}
\begin{subequations}\label{eq:Boussinesq}
    \begin{align}
        \frac{\partial\mathbf{u}}{\partial t} + (\mathbf{u} \cdot \nabla) \mathbf{u} + 2\mathbf{\Omega} \times \mathbf{u} &= -\nabla \Phi +b\mathbf{k}+\mathbf{\mathcal{F}_u} &\mbox{(momentum equation)} \label{eq:Boussinesq_mom} \\
        \nabla \cdot \mathbf{u} &= 0 &\mbox{(continuity equation)} \label{eq:Boussinesq_cont} \\
        \frac{\partial b}{\partial t} + \mathbf{u} \cdot \nabla b &= \mathcal{F}_b    &\mbox{(thermodynamic equation)} \label{eq:Boussinesq_therm}
    \end{align}
\end{subequations}
Here, $\mathbf{u}$ denotes the three-dimensional velocity, $\delta p$ the pressure perturbation associated with the density perturbation $\delta\rho$, $\Phi = \delta p / \rho_0$ the potential of the pressure-gradient accelerations, and $b =  -g \delta\rho / \rho_0$ the buoyancy ($g$ is an effective gravitational acceleration and $\mathbf{k}$ is the local vertical). The reference frame rotates with the constant angular velocity $\mathbf{\Omega}$ of the planetary rotation, as a result of which Coriolis accelerations ($2\mathbf{\Omega} \times \mathbf{u}$) appear in the momentum equation. Centrifugal accelerations are subsumed in $g$, the effective gravitational acceleration. The terms $\mathcal{F}_\mathbf{u}$ and $\mathcal{F}_b$ on the right-hand side represent dissipation and forcing terms (for example, friction and diabatic heating). The continuity equation reduces to the condition that the flow $\mathbf{u}$ is non-divergent. Sound waves are filtered from these equations because no time derivative appears in the continuity equation. In effect, the speed of sound is taken to be infinite, so that pressure adjusts instantaneously across the flow domain: it can be determined from a Poisson equation obtained by taking the divergence of the momentum equation and using the non-divergence condition to eliminate the time derivative.

To write the equations in a synthetic way, we introduce the state vector
\begin{equation}\label{eq:psi}
\mathbf{\Psi} = 
    \left(
        u\,,
        v\,,
        w\,,
        b 
    \right)^T 
\end{equation}
of the flow that contains all prognostic variables in Cartesian coordinates (the superscript $(\cdot)^{T}$ indicates the transpose). The set of equations (\ref{eq:Boussinesq}) can then be written compactly as
\begin{subequations}\label{eq:Boussinesq2}
    \begin{align}
        \frac{\partial\mathbf{\Psi}}{\partial t} + \nabla\cdot(\mathbf{\Psi} \otimes \mathbf{u}) &= \mathbf{L}\,\mathbf{\Psi} - \nabla \Phi + \mathbf{F} \\
        \nabla \cdot \mathbf{u} &= 0,
    \end{align}
\end{subequations}
where the outer product ($\otimes$) of the two vectors $\mathbf{\Psi}$ and $\mathbf{u}$ is defined as
\begin{equation}
    \mathbf{\Psi} \otimes \mathbf{u} = 
  \left[ {\begin{array}{ccc}
   \Psi_1 u_1 & \cdots & \Psi_1 u_3 \\
   \vdots & \ddots & \vdots \\
   \Psi_4 u_1 & \cdots & \Psi_4 u_3 
  \end{array} } \right]
 = (\Psi_i u_j)_{\substack{   
                            1 \leq i \leq 4 \\                                                                                    
                            1 \leq j \leq 3
                            }} \mbox{.}
\end{equation}
The components of the divergence of the second-order term $ \nabla\cdot (\mathbf{\Psi} \otimes \mathbf{u})$ are
\begin{equation}
  [\nabla\cdot(\mathbf{\Psi} \otimes \mathbf{u})]_i = \frac{\partial}{\partial r_j} \left[\Psi_i u_j\right] \mbox{,}
\end{equation}
where summation over repeated indices is implied. This notation extends the usual advection operator of a scalar field by a non-divergent flow to the advection of a vector field. The linear operator $\mathbf{L}$ contains accelerations owing to the Coriolis force, buoyancy, as well as nonconservative terms (e.g., friction or diabatic heating) that are linear in the state vector. The vector $\mathbf{F}$ contains all other nonconservative terms. We have expanded $\mathbf{u}$, the outer product and the $\nabla \cdot$ in Cartesian coordinates for simplicity, however the formalism is coordinate independent. 

Despite their relative simplicity, the Boussinesq equations are commonly used to study turbulence in boundary layers, in which density variations are weak. They also underlie classical conceptual models of large-scale atmospheric dynamics (e.g., quasigeostrophic models), in which they become quantitatively inaccurate but still qualitatively capture important atmospheric phenomena such as Rossby waves and baroclinic instability. The Boussinesq equations are well suited for our exposition of cumulant expansion approaches because they capture the essential nonlinearity of atmospheric flows: the conservative quadratic nonlinearity of the advection operators $(\mathbf{u} \cdot \nabla) \mathbf{u}$ and $\mathbf{u} \cdot \nabla b$. 

\subsection{Averaging operator}

Our interest is not in individual details of the atmospheric flows under consideration but in their statistics, including mean values and higher moments. Therefore, we define an averaging operator, denoted with an overline $\overline{(\cdot)}$, and decompose any scalar field $f(\mathbf{r},t)$ into a mean and a fluctuating part,
\begin{equation}\label{eq:decomp}
        f(\mathbf{r},t)=\bar f(\mathbf{r},t) + f'(\mathbf{r},t).
\end{equation}
The mean is in general a function of space and time. The fluctuating part is commonly referred to as an eddy. The averaging operator is taken to satisfy, for all scalar fields $f(\mathbf{r}, t)$ and $g(\mathbf{r}, t)$ and any constant $c$, the Reynolds properties \citep{monin1971}:
\begin{subequations}\label{eq:avg_pties}
    \begin{align}
       \overline{c} = c  & &  \\
       \overline{cf+g}=c\bar{f} + \bar{g}  & & \mbox{(linearity)} \\
       \overline{\bar{f}g} = \bar{f}\bar{g}  & &  \\
       \overline{\partial f}=\partial \bar{f} & & \mbox{(commutation with derivatives)}
    \end{align}
\end{subequations}
Properties (\ref{eq:avg_pties}a-c) imply that the averaging operator is a projection operator and so is idempotent $\bar{\bar{f}} = \bar f$. They make it possible to define the Reynolds decomposition
\begin{equation}\label{eq:reynolds}
        \overline{f g}= \bar{f}\bar{g} + \overline{f'g'}.
\end{equation}
For a vector quantity $\mathbf{\Psi}$, the average $\overline{\mathbf{\Psi}}$ is the component-wise average.

The choice of average is unspecified as long as it satisfies the Reynolds properties. Conceptually, ensemble averages are statistically meaningful, and they naturally satisfy the Reynolds properties. In practice, however, they can be difficult to obtain. Averages over sufficiently long times in statistically stationary (or slowly varying) flows or over sufficiently large regions in statistically homogeneous flows are more commonly used in practice, and also approximately satisfy the Reynolds properties. In concrete calculations, we choose the averages that are natural given the statistical symmetries of the problem under consideration. For example, in flows that are statistically invariant under translations along a spatial coordinate (e.g., along latitude circles), an average along that spatial coordinate suggests itself.

For more general flow equations with a variable density, the averaging operator above has to be replaced by a density-weighted average, to obtain consistent equations of motion for the statistical moments that resemble the Boussinesq moment equations formally. An example for the anelastic approximation is provided in \ref{a:anelastic}.

\subsection{Cumulant expansion}

\subsubsection*{First cumulant}

The first cumulant is the mean $\overline{\mathbf{\Psi}}(\mathbf{r},t)$, for which the equations of motion are obtained by averaging the equations of motion (\ref{eq:Boussinesq2}):
\begin{subequations}\label{eq:cum1}
    \begin{align}
        \frac{\partial\bar{\mathbf{\Psi}}}{\partial t} + 
        \nabla\cdot(\bar{\mathbf{\Psi}} \otimes \bar{\mathbf{u}}) &= 
        - \nabla\cdot(\overline{\mathbf{\Psi'} \otimes \mathbf{u'}}) + 
        \mathbf{L}\,\bar{\mathbf{\Psi}} - \nabla \bar \phi + \overline{\mathbf{F}}, \\
        \nabla \cdot \bar{\mathbf{u}} &= 0    \mbox{.}
    \end{align}
\end{subequations}
This involves the covariance $\nabla\cdot(\overline{\mathbf{\Psi'} \otimes \mathbf{u'}})$, which arises from the quadratic nonlinearity of the equations of motion. It represents eddy fluxes, for example, arising from advection of momentum fluctuations by the eddies (fluctuations) themselves. Because the equation for the mean involves a covariance, it is not closed.

\subsubsection*{Second cumulant}	
The second cumulant is the second central moment, or the covariance 
\begin{equation}
\mathbf{C}(\mathbf{r_1},\mathbf{r_2},t) = \overline{\mathbf{\Psi'}(\mathbf{r_1},t) \otimes \mathbf{\Psi'}(\mathbf{r_2},t)}.
\end{equation}
We only consider the prognostic variables here, because the diagnostic variables (and hence their statistics) can be obtained from them. We also only consider equal-time cumulants, that is, equal-time covariances between prognostic variables at the two points $\mathbf{r_1}$ and $\mathbf{r_2}$, which need not be equal. The covariance tensor is symmetric,
\begin{equation}\label{eq:symmetry}
    \mathbf{C}(\mathbf{r_1},\mathbf{r_2},t)=\mathbf{C}^T(\mathbf{r_2},\mathbf{r_1},t) \mbox{.}
\end{equation}
We additionally define the auxiliary covariances, 
\begin{subequations}\label{eq:extra_cum}
    \begin{align}
     \mathbf{C}^{\mathrm{\Phi}}(\mathbf{r_1},\mathbf{r_2},t)  &=  \overline{\Phi'(\mathbf{r_1},t)\mathbf{\Psi'}(\mathbf{r_2},t)} \\
     \mathbf{C}^u(\mathbf{r_1},\mathbf{r_2},t)  &=  \overline{\mathbf{\Psi'}(\mathbf{r_1},t) \otimes \mathbf{u'}(\mathbf{r_2},t)}  \mbox{.}
    \end{align}
\end{subequations}
The first, $\mathbf{C}^\Phi$, contains additional information about covariation of the prognostic fields $\mathbf{\Psi}$ with the pressure potential $\Phi$. These covariances can be calculated from $\mathbf{C}$ and $\overline{\Psi}$ because the pressure potential $\Phi$ is a diagnostic variable in the Boussinesq approximation. The second, $\mathbf{C}^{u}$, represents covariation of the velocity field with other prognostic variables and is already contained in $\mathbf{C}$. 

The second cumulant equation can be obtained from the equations of motion (\ref{eq:Boussinesq2}) by evaluating 
\begin{equation}
\frac{\partial}{\partial t}\mathbf{C}(\mathbf{r_1},\mathbf{r_2},t) = \overline{\mathbf{\Psi'}(\mathbf{r_1},t) \otimes \frac{\partial\mathbf{\Psi'}(\mathbf{r_2},t)}{\partial t}} + \overline{\frac{\partial\mathbf{\Psi'}(\mathbf{r_1},t)}{\partial t} \otimes \mathbf{\Psi'}(\mathbf{r_2},t)} \mbox{.}
\end{equation}
Discarding terms that are third order in fluctuating quantities, one obtains
    \begin{multline}       \label{eq:cum_tot_tensor}
        \frac{\partial}{\partial t}\mathbf{C}(\mathbf{r_1},\mathbf{r_2},t) 
        + \nabla_{\mathbf{r_1}} \cdot [\mathbf{C}(\mathbf{r_1},\mathbf{r_2},t) \otimes \bar{\mathbf{u}}(\mathbf{r_1},t)] = 
         - \mathbf{C}^u(\mathbf{r_1},\mathbf{r_2},t) \left[\nabla \bar{\Psi}(\mathbf{r_2},t)\right]^T \\
        + \mathbf{L}\, \mathbf{C}(\mathbf{r_1},\mathbf{r_2},t) 
        + \nabla_{\mathbf{r_1}} \mathbf{C^{\phi}}(\mathbf{r_1},\mathbf{r_2},t) 
        + \overline{\mathbf{F'(\mathbf{r_1},t)}\otimes\mathbf{\Psi'(\mathbf{r_2},t)}} 
        + \{\mathbf{r_1} \longleftrightarrow  \mathbf{r_2}\},
    \end{multline}
where $\{\mathbf{r_1} \longleftrightarrow  \mathbf{r_2}\}$ indicates the terms obtained by interchanging $\mathbf{r_1}$ and $\mathbf{r_2}$, which are necessary to ensure the symmetry (\ref{eq:symmetry}) of the covariance tensor. The third-order term $\mathbf{C}(\mathbf{r_1},\mathbf{r_2},t) \otimes \bar{\mathbf{u}}(\mathbf{r_1},t)$ is defined as in Cartesian coordinates, 
\begin{equation}
    \left[ \mathbf{C}(\mathbf{r_1},\mathbf{r_2},t) \otimes \bar{\mathbf{u}}(\mathbf{r_1},t) \right]_{p,q,i} = C_{p,q}(\mathbf{r_1},\mathbf{r_2},t) \bar u_i(\mathbf{r_1},t) \mbox{,}
\end{equation}
and its divergence by
\begin{equation}
    \left\{ \nabla_{\mathbf{r_1}} \cdot [\mathbf{C}(\mathbf{r_1},\mathbf{r_2},t) \otimes \bar{\mathbf{u}}(\mathbf{r_1},t)] \right\}_{p,q} = \frac{\partial}{\partial r_{1i}}\left[ C_{p,q} \bar u_i \right] \mbox{.}
\end{equation}
The flux $\mathbf{C}(\mathbf{r_1},\mathbf{r_2},t) \otimes \bar{\mathbf{u}}(\mathbf{r_1},t)$ represents the transport of spatial eddy correlations by the mean flow at $\mathbf{r_1}$. The term $-\mathbf{C}^u(\mathbf{r_1},\mathbf{r_2},t) \cdot \nabla \left[ \bar{\Psi}(\mathbf{r_2},t) \right]^{T}$ represents generation of covariance $\mathbf{C}(\mathbf{r_1},\mathbf{r_2},t)$ by advection down mean-flow gradients.

Additionally, continuity (\ref{eq:Boussinesq}b) implies that 
\begin{equation}
        \nabla_{\mathbf{r_2}} \cdot \mathbf{C}^u(\mathbf{r_1},\mathbf{r_2},t) = 
        \nabla_{\mathbf{r_1}} \cdot \mathbf{C}^u(\mathbf{r_2},\mathbf{r_1},t) = 0 \mbox{.}
\end{equation}
The set of equations for the first and second cumulants in this form is closed because the third cumulants, which would ordinarily appear in the second cumulant equations owing to the quadratic nonlinearity of the equations of motion, have been discarded. This second-order truncation of the otherwise infinite hierarchy of cumulant equations is referred to as a second-order cumulant expansion (CE2). In that CE2 assumes the first and second cumulants suffice to describe the flow statistics, it makes a normal approximation to the hierarchy of equations for the flow statistics. 

\subsubsection*{CE2 equations}

To summarize, the CE2 equations, with the second cumulant substituted in (\ref{eq:cum1}a), are given by
\begin{subequations}\label{eq:cum_tot_tensor}
    \begin{align}
        \frac{\partial\bar{\mathbf{\Psi}}(\mathbf{r},t)}{\partial t}  
        + \nabla\cdot\left[\bar{\mathbf{\Psi}}(\mathbf{r},t) \otimes \bar{\mathbf{u}}(\mathbf{r},t)\right] &= 
        - \nabla\cdot\mathbf{C}^u(\mathbf{r},\mathbf{r},t) \notag \\ 
        + \mathbf{L}\, \bar{\mathbf{\Psi}}(\mathbf{r},t) &- \nabla \bar \Phi(\mathbf{r},t) + \overline{\mathbf{F}}(\mathbf{r},t) \\
        \nabla \cdot \bar{\mathbf{u}} &= 0   \\       
        \frac{\partial}{\partial t}\mathbf{C}(\mathbf{r_1},\mathbf{r_2},t) 
        + \nabla_{\mathbf{r_1}} \cdot [\mathbf{C}(\mathbf{r_1},\mathbf{r_2},t) \otimes \bar{\mathbf{u}}(\mathbf{r_1},t)] &= 
        - \mathbf{C}^u(\mathbf{r_1},\mathbf{r_2},t) \left[\nabla \bar{\Psi}(\mathbf{r_2},t)\right]^T \notag \\
        + \mathbf{L} \mathbf{C}(\mathbf{r_1},\mathbf{r_2},t) 
        + \nabla_{\mathbf{r_1}} \mathbf{C^{\phi}}(\mathbf{r_1},\mathbf{r_2},t) 
        + &\overline{\mathbf{F'(\mathbf{r_1},t)}\otimes\mathbf{\Psi'(\mathbf{r_2},t)}} 
        + \{\mathbf{r_1} \longleftrightarrow  \mathbf{r_2}\} \label{eq:CE2_tot_C2} \\        
        \nabla_{\mathbf{r_2}} \cdot \mathbf{C}^u(\mathbf{r_1},\mathbf{r_2},t) & = 
        \nabla_{\mathbf{r_1}} \cdot \mathbf{C}^u(\mathbf{r_2},\mathbf{r_1},t) = 0 \mbox{.} \label{eq:CE2_tot_C2_nondiv}
    \end{align}
\end{subequations}
This set of equations involves  the 16 terms in $\mathbf{C}$ and the eight covariances $\mathbf{C}^\Phi(\mathbf{r_1},\mathbf{r_2},t)$ and $\mathbf{C}^\Phi(\mathbf{r_2},\mathbf{r_1},t)$. The 16 components of the second cumulant $\mathbf{C}$ are prognostic (evoloving according eq. \eqref{eq:CE2_tot_C2})), and the other 8 covariances components involve diagnostic correlations between the pressure potential and each of the other fields. The 8 corresponding diagnostic Poisson equations are obtained by taking the divergence with respect to $\mathbf{r_1}$  or $\mathbf{r_2}$ of the equation for $\mathbf{C}^u(\mathbf{r_1},\mathbf{r_2},t)$ that can be extracted from \eqref{eq:CE2_tot_C2}, and in which time tendencies vanish because of the non-divergence constraint \eqref{eq:CE2_tot_C2_nondiv}.

\subsubsection*{Physical properties}

CE2 is a realizable approximation, in the sense that the implied probability distribution functions are positive \citep{marston2014}.  The first cumulant equations (\ref{eq:cum_tot_tensor}a, b) are unchanged from the fully nonlinear system. Therefore, first-order invariants (mass and momentum) are conserved by the CE2 system in the absence of nonconservative effects. The third cumulants, which are neglected in the second cumulant equations (\ref{eq:cum_tot_tensor}c, d), redistribute second-order invariants (e.g., energy) among scales but do not generate or dissipate these invariants. Therefore, second-order invariants such as energy are likewise conserved by the CE2 system in the absence of nonconservative effects \citep[see][]{marston2014}.

\subsection{Quasi-linear approximation}

The CE2 equations can also be obtained by directly approximating the original equations of motion (\ref{eq:Boussinesq2}), making what has come to be called the quasi-linear (QL) approximation \citep{ogorman2007,srinivasan2012,Constantinou2014a,ait2014}. The QL approximation keeps nonlinear terms in the equation \eqref{eq:cum1} for the mean $\mathbf{\bar\Psi}$. However it linearizes the equation for the eddies $\mathbf{\Psi'}$, obtained by substracting (\ref{eq:cum1}a) from (\ref{eq:Boussinesq2}a),
\begin{equation}\label{eq:eddy_mean}
        \frac{\partial\mathbf{\Psi'}}{\partial t} 
        + \nabla\cdot(\mathbf{\bar\Psi} \otimes \mathbf{u'}) 
        + \nabla\cdot(\mathbf{\Psi'} \otimes \mathbf{\bar{u}}) 
        + \left[ \nabla\cdot(\mathbf{\Psi'} \otimes \mathbf{u'}) - \nabla\cdot(\overline{\mathbf{\Psi'} \otimes \mathbf{u'}}) \right] = 
        \mathbf{L}\,\mathbf{\Psi'} - \nabla \Phi + \mathbf{F'}\mbox{,}
\end{equation}
by setting
\begin{equation}
\nabla\cdot(\mathbf{\Psi'} \otimes \mathbf{u'}) = \nabla\cdot(\overline{\mathbf{\Psi'} \otimes \mathbf{u'}}).
\end{equation}
Hence, the QL approximation amounts to replacing in the Reynolds decomposition of the nonlinear term,
\begin{equation}
\mathbf\Psi \otimes \mathbf{u} = \Bar{\mathbf\Psi} \otimes \Bar{\mathbf{u}} + \Bar{\mathbf\Psi} \otimes \mathbf{u'} + \mathbf{\Psi'} \otimes \Bar{\mathbf{u}} + \mathbf{\Psi'} \otimes \mathbf{u'},
\end{equation}
the eddy--eddy interaction $\mathbf{\Psi'} \otimes \mathbf{u'}$ by its mean effect $\overline{\mathbf{\Psi'}\otimes \mathbf{u'}}$. Under the QL approximation, the Boussinesq equations (\ref{eq:Boussinesq2}) can then be written as 
\begin{subequations}\label{eq:QL}
    \begin{align}
        \frac{\partial{\mathbf\Psi}}{\partial t}  + \nabla \cdot \left( \mathbf\Psi \otimes \mathbf{u} \right) +  \nabla \cdot \left( \overline{\mathbf\Psi'\otimes \mathbf{u'}} -  \mathbf\Psi'\otimes \mathbf{u'}\right) &= \mathbf{L}\, \mathbf{\Psi} - \nabla \Phi+ \mathbf{F}, \\
        \nabla \cdot \mathbf{u} &= 0 \mbox{.}
    \end{align}
\end{subequations}
Because the QL equations retain as the only nonlinear interaction the interaction between eddies and mean fields, the corresponding cumulant equations are closed at second order, meaning no third-order terms appear in the second-order equations. The first two cumulant equations are exactly the CE2 equations (\ref{eq:cum_tot_tensor}). This makes it possible to simulate flows whose statistics satisfy the CE2 equations  (\ref{eq:cum_tot_tensor}) simply by simulating the QL equations (\ref{eq:QL}).

The QL truncation does not necessarily imply that eddy-eddy interactions and third-order correlations are equal to zero. 
However their evolution is decoupled from that of lower-order cumulants. This has to be kept in mind when interpreting instantaneous fields and statistics of flows simulated by the QL equations.  

\subsection{Differences between CE2 and S3T}

Stochastic Structural Stability Theory (S3T) is a second-order statistical approach to turbulent flows that is closely related to CE2. CE2 and S3T are sometimes presented as being equivalent in the literature because they share a similar mathematical formalism. 

However, CE2 and S3T differ in that S3T includes a small-scale stochastic forcing that is white in time and represents the scattering by missing eddy-eddy interactions. The resulting additional energy injection is balanced by large-scale linear damping. The stochastic forcing allows one to define rigorously an ensemble average over its realizations and permits a semi-analytical treatment of the second-order equations, whose solutions depend on the existence and statistics of the stochastic forcing. 

By contrast, CE2 uses the same forcing in the truncated equations as in the fully nonlinear equations, without attempting to parameterize eddy--eddy interactions. This choice is made for two reasons. First, a stochastic noise is not necessarily a realistic model for eddy-eddy interactions because these interactions do not inject energy but only redistribute it among scales. Second, forcing the flow at small scales might appear unnatural for a wide range of planetary flows, which are forced on larger scales. For example, large-scale motion in Earth's atmosphere is essentially driven by the planetary-scale radiative imbalance between the equator and the poles, rather than by energy injection at small scales. 

Another difference is that S3T is generally applied to flows for which the linear operator representing eddy--mean flow interactions does not have unstable modes (the mathematical development of S3T relies on non-normal growth and decay of stable modes, excited by the stochastic noise). 
In this framework, S3T has been used in barotropic rotating flows to study instabilities of zonal flows \citep[][]{bakas2013,parker2014} and the dynamics of zonal \citep{farrell2003,farrell2007,constantinou2014b} and non-zonal \citep{bakas2014} coherent structures. Nevertheless, the simulation of unstable flows at second order is possible and well-defined mathematically. Hence, CE2 (or its QL analogue) has been applied to unstable flows in barotropic \citep{marston2008} and baroclinic \citep[][]{ogorman2007,ait2014} settings. Evaluating the ability of CE2 in capturing unstable flows is at the cornerstone of the present paper. 

\section{Dry convective boundary layer}\label{sec:boundary_layer}

The dynamics of boundary layers and clouds involve flow scales of order meters to a few kilometers. In addition, condensate formation and evaporation take place at microscales. By contrast, typical climate models have a horizontal resolution of order 100~km. Therefore, the dynamics of boundary layers and clouds are subgrid-scale processes in climate models, which must be represented parametrically in terms of the resolved large-scale dynamics \citep[e.g.,][]{beljaars92a,smith1997}. Uncertainties about these parameterization schemes are the dominant contributor to uncertainties in climate change projections \citep[e.g.,][]{stephens2005,bony2006,soden2006,webb2013,stevens2013,vial2013,brient2016}.

Most current parameterization schemes for the dynamics of clouds and boundary layers truncate the hierarchy of moment (or cumulant) equations at first order and represent the second-order subgrid-scale fluxes appearing in the first-order equations semi-empirically. For example, turbulent fluxes in boundary layers are often represented as diffusive fluxes down mean gradients, with diffusivities estimated from approximate spatially local second-order equations for turbulence kinetic energy \citep[e.g.,][]{mellor82a,beljaars92a}. Turbulent fluxes in convective clouds, on the other hand, are often represented as vertically non-local entraining plumes that extend deep into the boundary layer. The parameters determining the mass fluxes in the plumes are estimated from energy equations \citep[e.g.,][]{arakawa74,gregory97b}.

CE2 may offer a path toward improved and unified representations of subgrid-scale turbulent fluxes (e.g. in boundary layers and in convective clouds) in climate models, because CE2 explicitly retains spatial nonlocality and interactions between fluctuations (plumes or eddies) and the environment (mean field). Here we compare a QL simulation and a fully nonlinear simulation of the simplest convective boundary layer, a dry convective boundary layer, and demonstrate the potential and limitations of CE2 to represent the statistics of its dynamics. We conducted a large-eddy simulation (LES) of a dry convective boundary layer growing into a stable background stratification as a heat flux is imposed at the surface \citep{soares04}. We then compared this fully nonlinear simulation with a simulation in which the equations were replaced by the corresponding QL equations (\ref{eq:QL}).

\subsection{Large-eddy simulations}
	
	\subsubsection*{Setup}
	
The LES code solves the anelastic equations and is described in \citet{pressel2015}. Like the Boussinesq approximation, the anelastic equations are non-hydrostatic and filter sound waves by neglecting dynamic density variations except where they affect buoyancy. In contrast to the Boussinesq approximation, the background state depends on the vertical coordinate. Hence, the anelastic equations can capture the dynamics of flows with substantial stratification and so  are better suited to study atmospheric convection. The anelastic approximation and its cumulant expansion are described in \ref{a:anelastic}. 

Our anelastic LES code uses the specific entropy $s$ and the three-dimensional velocity field $\mathbf{u}$ as prognostic variables \citep{pauluis2008}. We use a second-order central difference spatial discretization scheme with strongly stability preserving Runge-Kutta timestepping \citep{spiteri2002}. The time step is dynamically adjusted to ensure a Courant number near 0.3. Because LES merely resolves the most energetic scales of the flow, subgrid-scale (SGS) motions must also be modeled, and we do so by applying a constant eddy diffusivity of $\nu = 1.2\,\unit{m^2\,s^{-1}}$ throughout the domain. We choose a constant diffusivity to avoid the nonlinearities that appear in the computation of the eddy viscosity by more advanced SGS schemes such as the Smagorinsky-Lilly closure \citep{smagorinsky1963,lilly1962}, which would need to be linearized in a QL simulation; constant diffusion as an SGS closure allows a more direct comparison of fully nonlinear and QL simulations, notwithstanding that it is an inferior SGS closure. The domain extends $6.4\unit{km}\times 6.4\unit{km}$ in the horizontal and $3.75\unit{km}$ in the vertical, with a horizontal and vertical resolution of $25\unit{m}$. Horizontal boundary conditions are doubly periodic. At the upper boundary, flow fields are linearly relaxed over a $800\unit{m}$ deep layer toward the horizontal mean flow, which is almost motionless (horizontal mean velocities are of the order $0.1 \unit{m} \unit{s}^{-1}$). The relaxation coefficient varies from $\tau = 0$ at the bottom of the layer to $\tau = (100\unit{s})^{-1}$ at the top. 

The initial state is stably stratified with a potential temperature $\theta$ that increases linearly with height, at a rate of $2\,\unit{K\,km^{-1}}$. Here, the potential temperature is related to the specific entropy by $\theta = \tilde T \exp[(s-\tilde s)/c_p]$, where $c_p$ is the specific heat at constant pressure, $\tilde T$ is a standard temperature, and $\tilde s$ is a standard specific entropy at the standard temperature $\tilde T$ and standard pressure $\tilde p = 1000~\mathrm{hPa}$.\footnote{Using the ideal-gas law, it can be verified that $\theta$ is  the usual potential temperature with reference pressure $\tilde p$: $\theta = T(\tilde p/p_0)^{R/c_p}$, with specific gas constant $R$. However, this potential temperature is evaluated at the anelastic reference-state pressure $p_0(z)$ rather than at the in situ pressure $p$, as is required for thermodynamic consistency of the anelastic approximation \citep{pauluis2008}.} The initial state is destabilized by imposing a constant sensible heat (enthalpy) flux  of $70.46\,\unit{W\,m^{-2}}$ at the lower boundary. Normally distributed random fluctuations of the potential temperature with amplitude $0.1\unit{K}$ in the lowest $200\unit{m}$ break the horizontal homogeneity of the initial state and allow the generation of turbulent motions. The initial flow is uniform, horizontal, and has a speed of $0.01 \unit{m} \unit{s}^{-1}$. Together with the drag at the lower boundary, this allows for turbulent momentum fluxes to develop \citep{soares04}. We ran simulations for up to 12 simulated hours, over which a dry convective boundary layer forms and grows as a result of the heating at the bottom.

Because of the statistical horizontal homogeneity of the flow, we use the horizontal average to define mean fields and eddies, as in Eq.~(\ref{eq:decomp}). The QL truncation (\ref{eq:QL}), here with the state vector $\mathbf\Psi = \left(u,v,w,s\right)^T$, is implemented by removing at every time step the nonlinear eddy--eddy interactions from the tendencies of all prognostic variables.\footnote{The QL truncation (\ref{eq:QL}) is valid for the Boussinesq approximation. The anelastic approximation requires us to use an average weighted by the background density, as explained in \ref{a:anelastic}.  But because averages are here performed on horizontal surfaces with constant background density, the QL approximation for the anelastic system is also given by (\ref{eq:QL}).} \citet{herring1963} similarly used the QL approximation to study thermal convection between two parallel horizontal plates.

\subsection{Results}
	
\begin{figure}[ht]
  \noindent
  \includegraphics[width=\textwidth]{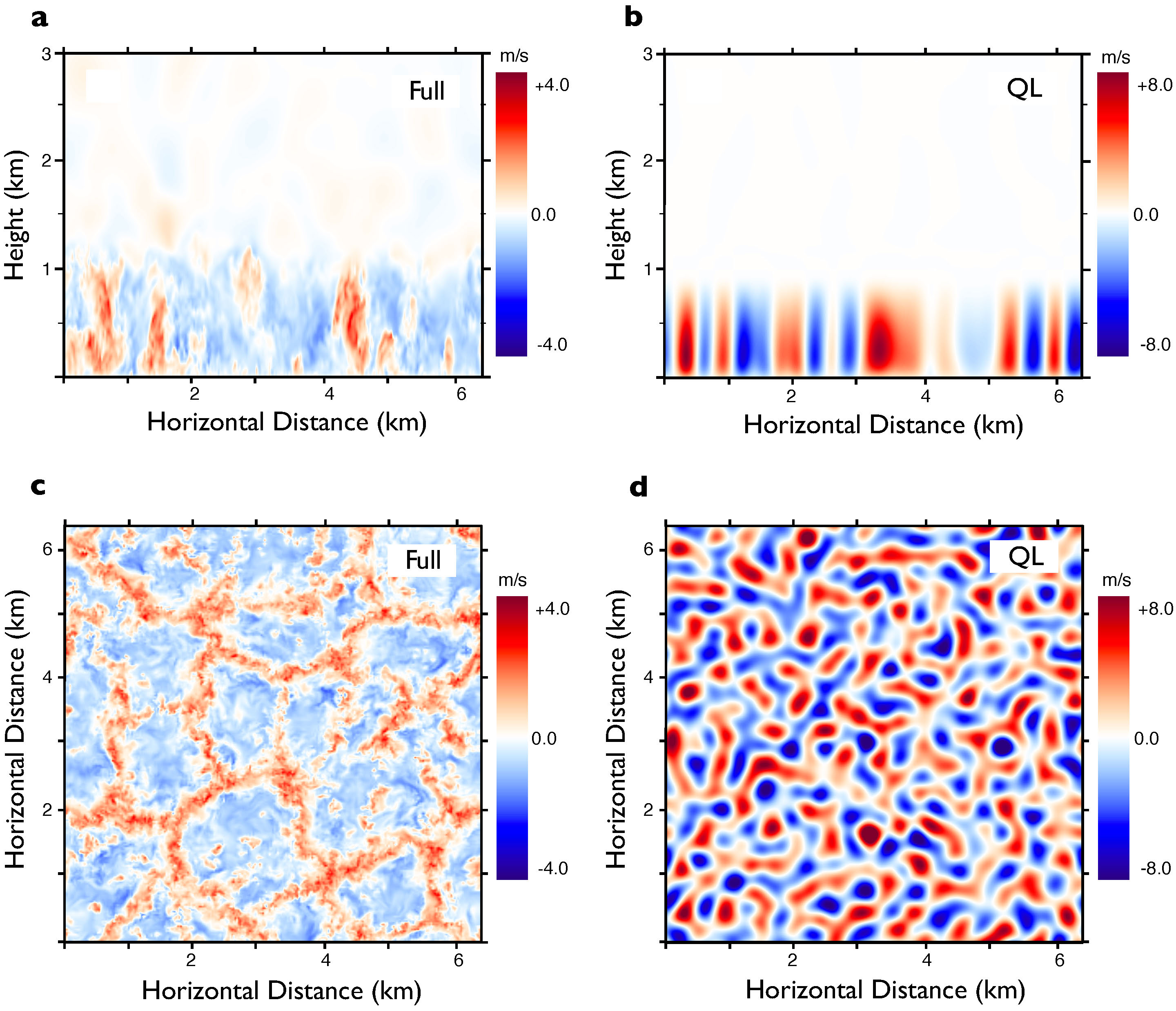}\\
  \caption{Instantaneous cross sections of the vertical velocity after 4 hours. Vertical cross sections in (a) fully nonlinear simulation and (b) QL simulation. Horizontal cross sections at 250~m altitude in (c) fully nonlinear simulation and (d) QL simulation. Note the different color scales for the fully nonlinear and QL simulation. }\label{fig:Maps_DCBL}
\end{figure}

Figure~\ref{fig:Maps_DCBL} compares vertical and horizontal cross sections of the vertical velocity field in the fully nonlinear and in the QL simulations. The vertical cross sections (Fig.~\ref{fig:Maps_DCBL}a,b) show that upward motion mainly occurs in vertically coherent updrafts, as is well known \citep[e.g.,][]{schmidt1989}. In the QL simulation, updrafts are more coherent because small-scale structures are missing while the larger scales are well represented. The horizontal cross section in the fully nonlinear simulation reveals that the updrafts are organized into polygonal convective cells (Fig.~\ref{fig:Maps_DCBL}c). Such cellular patterns are well known to arise near the onset of thermal instability of a fluid heated from below \citep[][chapter~2]{chandrasekhar1961}. The QL simulation does not capture these horizontal correlation patterns (Fig.~\ref{fig:Maps_DCBL}d), probably because eddy-eddy interactions in the horizontal plane play a role in generating the smaller scales of the cellular patterns.  The vertical velocity fluctuations $w'$ are distributed more symmetrically around zero in the QL simulation: updrafts and downdrafts are of similar size and strength (Fig.~\ref{fig:Maps_DCBL}b, d). This stands in contrast to the fully nonlinear simulations, in which updrafts are faster and narrower and downdrafts slower and broader. 

\begin{figure}[ht]
  \noindent\includegraphics[width=\textwidth]{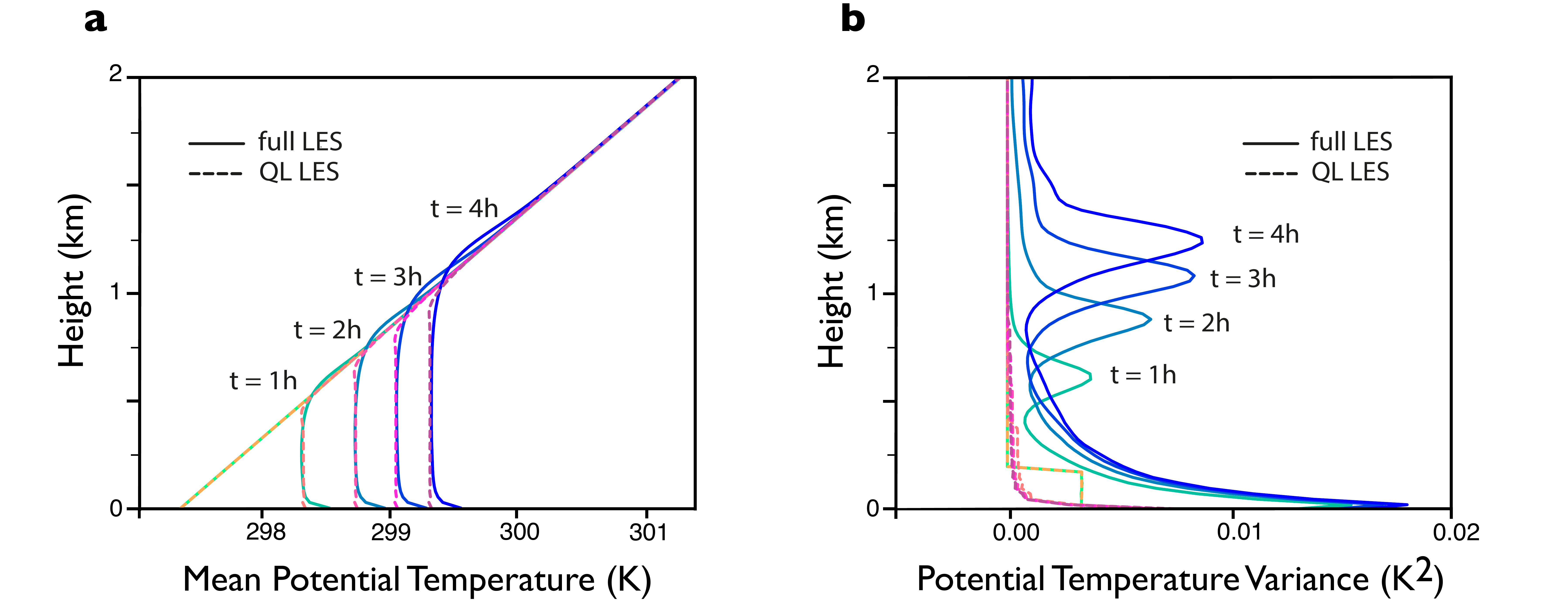}\\
  \caption{(a) Vertical profiles of potential temperature at indicated times in the fully nonlinear (solid lines) and QL (dashed lines) simulations. (b) Corresponding vertical profiles of the potential temperature variance $\overline{\theta'^2}$. }\label{fig:profils_DCBL}
\end{figure}

\begin{figure}[ht]
  \noindent\includegraphics[width=\textwidth]{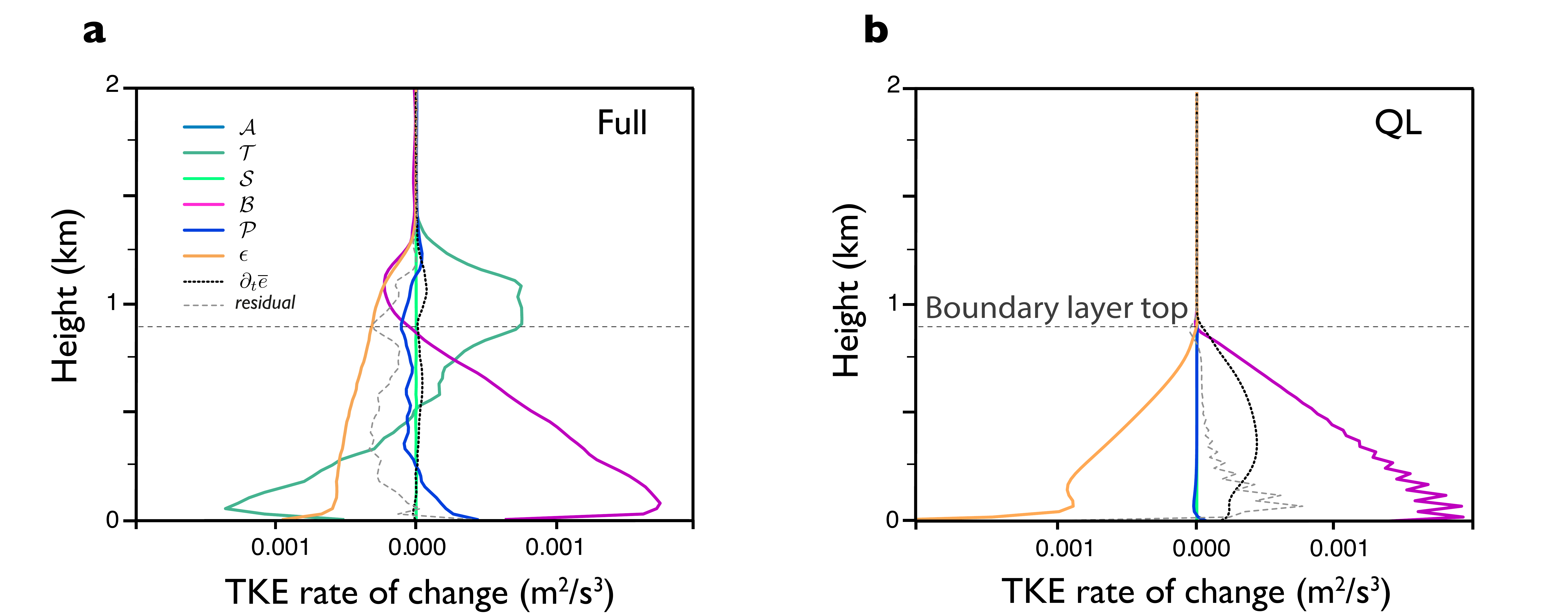}\\
  \caption{Turbulence kinetic energy  budget of the fully nonlinear (a) and QL simulations (b). The different terms denote the TKE advection $\mathcal A$, TKE transport by eddies $\mathcal T$, shear production $\mathcal S$, buoyancy production $\mathcal B$, pressure correlation term $\mathcal P$, dissipation to subgrid-scales $\mathcal D$ and rate of change of TKE $\mathcal R$. Also shown are the residuals, which are of the same order as the SGS diffusion term in the fully nonlinear simulation \citep[cf.][]{heinze2015} but are smaller in the QL simulation, likely because the latter is smoother, thus limiting numerical diffusion.}\label{fig:plots_TKE}
\end{figure}

Figure~\ref{fig:profils_DCBL}a shows the evolution of the vertical profile of the potential temperature in the full and QL simulations. Because of the constant heating at the lower boundary and the absence of any thermal relaxation in the fluid interior, the boundary layer continually deepens and does not reach a statistically steady state (Fig.~\ref{fig:profils_DCBL}a). The boundary layer is well mixed with homogenized potential temperature below its top. This indicates that the BL is close to the critical state for thermal instability. The QL simulation captures quite accurately the growth rate of the boundary layer and the mixing of potential temperature below its top. This is because the growth of the boundary layer mainly arises through interactions of fluctuations with the mean flow, which are retained in the QL simulation. 

Above the top of the boundary layer, defined as the altitude below which potential temperature is well mixed, the QL and the fully nonlinear simulations exhibit important differences.  In the QL simulation, the mean potential temperature profile is identical to the initial profile. By contrast, the fully nonlinear simulation shows a layer of strong stability associated with convective overshoots of thermal updrafts into the free atmosphere and the downward entrainment of warmer free-atmospheric air into the boundary layer \citep[e.g.,][]{De-Roode04a}. These overshoots are missing in the QL simulation, as they are in first-order diffusive closure schemes for convective boundary layers \citep[e.g.,][]{stull1988}. The difference at the top of the boundary layer is emphasized in the profile of temperature variance (Fig.~\ref{fig:profils_DCBL}b), which shows a strong peak above the top of the boundary layer for the fully nonlinear simulation but vanishing variance throughout the mixed layer and aloft for the QL simulation. Near the surface, the mean potential temperature is reduced in the QL simulation, which reflects a more efficient upward transport of heat into the interior of the boundary layer. This probably can be ascribed to the more intense vertical velocities and the strengthened correlations between buoyancy and vertical velocity fluctuations in the more coherent updrafts of the QL flow. 

An inability of diffusive closures and the QL simulation to represent the turbulent transport of turbulence kinetic energy (TKE) lies behind the failure of the QL simulation and diffusive closures to represent convective overshoots. We take TKE to be the kinetic energy of the resolved scales of the LES: $e = 0.5\,\left(u'^2 + v'^2 + w'^2\right)$. Its mean budget is derived from the momentum equations and reads 
\begin{multline}       \label{eq:TKE}
\partial_t \overline{e} =  - \underbrace{\overline{w}\, \partial_z \overline e}_{\mathcal A} 
- \underbrace{\frac{1}{\rho_0}\partial_z \left( \rho_0 \overline{w'e}\right)}_{\mathcal T} -  \underbrace{\left[\overline{w'u'}\partial_z\overline{u} + \overline{w'v'}\partial_z\overline{v} +\overline{w'w'}\partial_z \overline{w}\right]}_{\mathcal S}\\
+ \underbrace{\overline{w'b'}  }_{\mathcal B}
- \underbrace{\frac{1}{\rho_0}\partial_z\left(\overline{w'p'}\right)}_{\mathcal P} + \epsilon.
    \end{multline}
The right-hand side contains all terms that generate, destroy, or redistribute TKE: advection by the mean flow $\mathcal A$ and the eddies $\mathcal T$, shear production $\mathcal S$, buoyancy production $\mathcal B$ ($b' = g\theta'/\theta_0$ denotes buoyancy fluctuations), the pressure correlation term $\mathcal P$ and dissipation $\epsilon$. The dissipation $\epsilon$ denotes the energy flux from the resolved scales to sub-grid scales, which in our case is parameterized as viscous dissipation of the resolved fields with constant eddy viscosity $\nu$. All but the turbulent transport term $\mathcal T = -1/\rho_0 \partial_z \left( \rho_0 \overline{w'e}\right)$ are of second order in fluctuations and hence are retained in a QL truncation.

The TKE budgets for the fully nonlinear and QL simulations are shown in Figure~\ref{fig:plots_TKE}. The dominance of the buoyancy production term relative to the shear production term in both plots indicates that the flow is thermally driven. In the fully nonlinear simulation, buoyancy production is positive throughout the well-mixed part of the boundary layer (i.e., upward buoyancy flux), zero at the top of the boundary layer, and negative aloft (i.e., downward buoyancy flux). The negative flux is related to the overshooting thermals, which trigger a downward entrainment flux of free atmospheric air into the boundary layer. This negative flux is missing in the QL simulation, in which the buoyancy flux is zero above the top of the boundary layer. However, the buoyancy flux is well captured in the interior of the boundary layer.

The triple correlation transport term $\mathcal T$ represents the vertical transport of TKE by eddies. In the fully nonlinear simulation, eddies transport TKE from lower levels with high TKE to higher levels with low TKE. This transport seems to be crucial for the growth of TKE in the upper part of the boundary layer and especially across the top of the boundary layer. The neglect of this term in the QL dynamics is responsible for the missing overshoots of thermals across the boundary layer top and the missing associated negative buoyancy flux. The transport term $\mathcal T$ can be evaluated in the QL simulations and actually is nonzero. However, it decouples from the second-order dynamics and so does not affect the TKE budget.

\subsection{Implications}
	
These results show that a QL simulation can capture important aspects of the evolution of a dry convective boundary layer, such as its well-mixed nature and its growth over time. They suggest that a corresponding CE2 closure would also capture the relevant first-order statistics and, therefore, has promise as a nonlocal second-order closure in climate models. Deficiencies such as the missing convective overshoots at the top of the boundary layer may be remedied, for example, by adding a linear, diffusive transport term in the prognostic equations of the second-order moments \citep[e.g.][]{mellor73aa}. The resulting parameterization scheme would solve directly for the 1st- and 2nd-order statistics in every grid box of the large-scale model. 

The applicability of such a scheme depends on whether the corresponding CE2 equations can be solved efficiently---much more efficiently than an explicit QL simulation can be run. This will require a simplified representation of horizontal covariances, for example, by assuming approximate statistical symmetries such as horizontal homogeneity and isotropy of fluctuations. But the important nonlocal effects of vertical covariances need to be retained.

The eddy--eddy interactions neglected in the QL simulations may be even more critical for moist convection than they are for the dry convective boundary layer \citep{Firl2015}. Hence, more sophisticated approaches may be needed to expand the field of application of CE2 closure schemes to moist boundary layer dynamics. Exploring to what degree the CE2 approximation and its extensions can capture the dynamics of clouds and moist boundary layers promises to be a fruitful area of study. 

\section{Large-scale eddy decay on the rotating sphere}\label{sec:barotropic}

While the preceding example concerned the applicability of CE2 to atmospheric dynamics on scales of meters to kilometers, we now turn to a prototype problem for atmospheric dynamics on scales of hundreds to thousands of kilometers. Eddies on such large scales are essentially the well-known weather systems. They are generated by baroclinic instability and are fundamental for maintaining Earth's climate because they are responsible for the bulk of the atmospheric transport of energy, water vapor, and angular momentum. Through these transports, they shape the distribution of temperature, precipitation, and winds at the surface \citep{peixoto1992}. The fundamental balances governing such large-scale eddies are different than those in the boundary layer. The Coriolis force due to the planetary rotation and the average stable stratification become of primary importance, leading to flows that are more two-dimensional in character than boundary-layer flows. A two-dimensional (latitude-longitude) model suffices to illustrate some of the issues one encounters if one wants to develop a closure for the large-scale dynamics of the atmosphere based on cumulant expansions. 

\subsection{Barotropic model for the upper troposphere}

Large-scale eddies in Earth's atmosphere are generated near the surface in midlatitudes, propagate upward through the troposphere, and propagate meridionally in the upper troposphere \citep{simmons1978,edmon1980,heldhoskins1985,thorncroft1993}. Their meridional transport and eventual dissipation by wave breaking in latitude bands away from their generation latitudes is what gives rise to their meridional angular momentum transport: Large-scale eddies in rapidly rotating atmospheres transport angular momentum from their dissipation latitudes into their generation latitudes, that is, in the opposite direction to their meridional propagation \citep{kuo1951,held1975,held1999b}. This angular momentum transport ultimately shapes the strength and distribution of surface winds, with easterlies in the tropics, westerlies in midlatitudes, and weak easterlies again in polar latitudes \citep[see][for a review]{schneider2006b}. To understand the strength and distribution of surface winds, it is therefore necessary to understand the meridional propagation and dissipation of large-scale eddies, which are concentrated in the upper troposphere \citep{ait2014}. The simplest model that captures these processes is the barotropic model---a model of a two-dimensional fluid layer on a sphere, thought to represent a layer in the upper troposphere \citep[e.g.,][]{held1987}.
    
\subsubsection*{Equations of motion}

The equations governing barotropic flow can be derived from the Boussinesq equations (\ref{eq:Boussinesq}). Consider a Boussinesq flow on a sphere of radius $a$ rotating at constant spin angular velocity $\mathbf{\Omega}$. We further assume that the flow is two-dimensional on the sphere: $\mathbf{u}\cdot\mathbf{e}_r=0$, with $\mathbf{e}_{r}$ being the radial coordinate. In that case, the momentum and continuity equations (\ref{eq:Boussinesq_mom},~\ref{eq:Boussinesq_cont}) reduce to 
\begin{subequations}\label{eq:barotropic}
    \begin{align}
    \frac{\partial \mathbf{v}}{\partial t} + \mathbf{v}\cdot\nabla \mathbf{v} + 2\mathbf{\Omega} \times \mathbf{v} &=         -\nabla \Phi + \mathcal{F}_\mathbf{v} \label{eq:barotropic_momentum}\\
    \nabla \cdot \mathbf{v} &= 0 \mbox{.}
    \end{align}
\end{subequations}
The horizontal components of the wind and of the forcing are denoted $\mathbf{v}$ and $\mathcal{F}_\mathbf{v}$. Because the flow $\mathbf{v}$ is two-dimensional on the sphere, only the local vertical component $\Omega \sin\phi\,\mathbf{e}_r$ (latitude $\phi$) of $\mathbf{\Omega}$ yields non-zero terms in \eqref{eq:barotropic_momentum}. Taking the curl of the momentum equation and projecting it onto the radial direction $\mathbf{e}_r$ yields the two-dimensional barotropic vorticity equation,
\begin{equation}\label{eq:barotropic_pv}
    \frac{\partial q}{\partial t} + \mathbf{v}\cdot\nabla q =(\nabla \times \mathcal{F}_{\mathbf{v}}) \cdot \mathbf{e}_r 
\mbox{.}
\end{equation}
The flow is entirely described by this equation for the absolute vorticity 
\begin{equation}\label{eq:pv}
    q = f + \zeta,
\end{equation}
where the Coriolis parameter $f(\phi) = 2\Omega\,\sin\phi$ represents the vorticity of solid body rotation, and $\zeta = (\nabla \times \mathbf{v})\cdot \mathbf{e}_{r}$ is the relative vorticity in the radial direction $\mathbf{e}_{r}$, relative to the rotating reference frame. The advection term $\mathbf{v}\cdot\nabla q$ contains the advection of planetary vorticity $\mathbf{v} \cdot \nabla f=\beta v$, with $\beta = a^{-1} \partial_\phi f = 2\Omega a^{-1} \cos\phi$, which arises from the curl of the Coriolis force. This term is commonly referred to as the $\beta$-term. The vorticity equation (\ref{eq:barotropic_pv}) contains the entire dynamics of the flow because an incompressible two-dimensional flow is described by a streamfunction $\psi$, defined such that 
\begin{subequations}\label{eq:barotropic_stream}
    \begin{align}
        \mathbf{v}  &= \nabla \times (\psi\, \mathbf{e}_{r}) \mbox{,} \label{eq:baro_u} \\
        \zeta  &= \nabla^2\psi \mbox{.} \label{eq:baro_vort}
    \end{align}
\end{subequations}
That is, if the relative vorticity $\zeta$ is known, the advecting velocity \eqref{eq:baro_u} can be determined from the streamfunction, which is the solution of a Poisson equation \eqref{eq:baro_vort}. Thus, the equations of motion \eqref{eq:barotropic_pv} and \eqref{eq:barotropic_stream}, supplemented by appropriate boundary conditions, specify the dynamics completely. 

The equations of motion (\ref{eq:barotropic_pv})-(\ref{eq:barotropic_stream}) can be nondimensionalised using the planetary radius $a$ as the typical length scale and the length of the day $2\pi\Omega^{-1}$ as the typical time scale. With that nondimensionalisation, the operators $\nabla$, $\nabla \times$ and $\nabla^2$ become operators on the unit sphere, and the angular velocity $\Omega$ becomes $2\pi$. Throughout the rest of the paper, we will use there nondimensionalized quantities, unless otherwise specified.

\subsubsection*{Eddy--mean flow decomposition} 

We consider situations in which the boundary conditions of the problem are statistically symmetric under rotations around the planet's spin axis, so that the flow statistics (but not the instantaneous flow itself) can be expected to be axisymmetric. A zonal average $\overline{(\cdot)}$ then suggests itself. Decomposing flow fields in the nondimensional barotropic vorticity equation into zonal means $\overline{(\cdot)}$ and eddies $(\cdot)' = (\cdot) - \overline{(\cdot)}$ yields the mean and eddy vorticity equations,
\begin{subequations}\label{eq:vort_eddy_mean}
\begin{align}
    \frac{\partial\bar\zeta}{\partial t} &= -\overline{J_n(\psi',\zeta')}, \label{eq:vort_eddy_mean_1} \\
    \left(\frac{\partial}{\partial t} - \frac{\bar{u}}{\cos\phi}\frac{\partial}{\partial \lambda} \right)\zeta' &= -\left[J_n(\psi',\zeta') - \overline{J_n(\psi',\zeta')}\right] - v'\left(\beta + \frac{\partial \bar\zeta}{\partial\phi}\right)\mbox{.} \label{eq:vort_eddy_mean_2}
\end{align}
\end{subequations}
Here $\beta = 2\Omega cos\phi$, with $\Omega = 2\pi$, is the nondimensional meridional derivative of the  Coriolis parameter, and the Jacobian
\begin{equation}
J_n(\psi,\zeta) = \frac{1}{\cos\phi} \left( \frac{\partial \psi}{\partial \lambda} \frac{\partial \zeta}{\partial \phi} -  \frac{\partial \psi}{\partial \phi} \frac{\partial \zeta}{\partial \lambda} \right)
\end{equation}
represents the advection on the unit sphere of vorticity $\zeta$ by the zonal ($u$) and meridional ($v$) flow implied by the streamfunction $\psi$:
\begin{equation}
u = -\frac{\partial \psi}{\partial \phi}, \qquad
v =  \frac{1}{\cos\phi}\frac{\partial \psi}{\partial \lambda}.
 \end{equation}
The streamfunction-vorticity relations are
\begin{subequations}
    \begin{align}
        \overline{\zeta}  &= \nabla_n^2\bar\psi, \\ 
        \zeta'  &= \nabla_n^2\psi', 
    \end{align}
\end{subequations}
with $\nabla_n$ likewise defined on the unit sphere. 

Equation \eqref{eq:vort_eddy_mean} is essentially \eqref{eq:eddy_mean} for the 2D barotropic vorticity equation. The mean flow evolves in time due to vorticity fluxes $-\overline{J_n(\psi',\zeta')}$. The eddy vorticity budget involves shear by the mean flow $\bar{u}/\cos\phi \, \partial_{\lambda} \zeta'$, eddy--eddy interactions $\left[J_n(\psi',\zeta') - \overline{J_n(\psi',\zeta')}\right]$, the $\beta$-term  $\beta v'$ and the advection of mean--flow vorticity $\partial_{\phi} \bar\zeta\, v'$.

\subsubsection*{Cumulants}

The first cumulant is the mean vorticity $\bar\zeta(\phi,t)$, and the second cumulant is the vorticity equal-time two-point correlation:
\begin{subequations}\label{eq:def_cumulants_barotropic}
    \begin{align}
        \bar\zeta(\mathbf{r},t) &= \bar\zeta(\phi,t) ,  \\
        c(\mathbf{r_1},\mathbf{r_2},t) &=\overline{\zeta'(\mathbf{r_1},t)\,\zeta'(\mathbf{r_2},t)}  = C(\phi_1,\phi_2,\lambda_1-\lambda_2, t).  &
    \end{align}
\end{subequations}
The first cumulant depends on the latitude $\phi$, and the second cumulant depends on the latitudes $\phi_1$ and $\phi_2$ and, because of the statistical axisymmetry of the equations, on the longitude difference $\lambda_1-\lambda_2$ \citep{marston2008}. Because the flow is entirely determined by the scalar $q$ (or $\zeta$), all other correlations are determined by the scalar cumulant $c$. Hence, moments like $\overline{\zeta'(\mathbf{r_1},t) \otimes \mathbf{u}'(\mathbf{r_2},t)}$ or $\overline{\mathbf{u}'(\mathbf{r_1},t) \otimes \mathbf{u}'(\mathbf{r_2},t)}$ can be calculated from $\bar\zeta$ and $c$ \citep[e.g.,][]{marston2008,srinivasan2012,marston2014}. 

The CE2 equations for this problem are given in \cite{marston2008,marston2014}. They are of the form (\ref{eq:cum_tot_tensor}a) and (\ref{eq:cum_tot_tensor}c), with vorticity fluxes appearing as the essential eddy terms. The equivalent QL system is \eqref{eq:vort_eddy_mean} where the eddy-eddy interactions are neglected. 

\subsubsection*{Numerical implementation}

We simulate a barotropic flow on a sphere, specified by the equations of motion \eqref{eq:barotropic_pv} and \eqref{eq:barotropic_stream} on a spherical geodesic grid \citep{heikesrandall1995a,heikesrandall1995b,qimarston2014} with $163,842$ cells. To remove enstrophy that cascades to the smallest scales, hyperviscous dissipation $\nu (\nabla^2+2) \nabla^6 \zeta$ is included, where the term $(\nabla^2 + 2)$ ensures that angular momentum is conserved.   The hyperviscosity coefficient $\nu$ is chosen such that the smallest resolved mode decays with an $e$-folding time of 5. The vorticity is stepped forward in time by a second-order leapfrog scheme using the Robert-Asselin-Williams filter \citep{williams2009}.  The time step is fixed at $\Delta t = 0.01$.

Explicit time integration of the cumulant equations is carried out in spectral space using a 4th-order Runge-Kutta algorithm with an adaptive time step. We adopt the spectral truncation $0 \leq \ell \leq L$ on spherical wavenumber $\ell$, with the zonal wavenumbers restricted to $|m| \leq \min\{\ell, M\}$.  We choose spectral cutoffs $L = 60$ and $M = 30$. Hyperviscosity is adjusted to ensure the same $e$-folding time at the maximum wavenumber $\ell = L$ as on the smallest resolved spatial scales on the geodesic grid.

To verify that the spectral cumulant simulation has sufficient resolution and can be compared to the geodesic grid model, a simulation of the fluid is also performed in a pure spectral calculation with the same numerical methods and resolution as for the cumulant equations. The agreement between the spectral and geodesic models is excellent.  QL simulations are performed in spectral space by removing the triads that correspond to the interaction of two eddies, each with non-zero zonal wavenumber.

A program that implements fully nonlinear simulations on the spherical geodesic grid, and the nonlinear, QL, and CE2 simulations in spectral space, is freely available.\footnote{The application ``GCM'' is available for OS X 10.9 and higher on the Apple Mac App Store at URL http://appstore.com/mac/gcm}  More details about the simulations and the cumulant expansions can be found in \cite{marston2014}.

\subsection{Eddy lifecycle simulations} 

\subsubsection*{Setup} 

To illustrate situations when CE2 and QL approaches succeed or fail at capturing barotropic flow dynamics, we simulate the evolution of an initial zonal flow $U(\phi)$ with a superimposed initial disturbance (eddy) with vorticity $\zeta_i(\phi,\lambda)$. The zonal flow $U$ and disturbance $\zeta_i$ mimic the upper-tropospheric jet stream and disturbances that may originate, for example, from lower-tropospheric baroclinic instability. The setup is inspired by \cite{held1987} and uses
\begin{subequations}\label{eq:init}
\begin{align}
U(\phi) & = A\cos\phi - B\cos^2\phi + C\sin^2\phi\cos^4\phi - D \cos^6\phi, \\ 
\zeta_i(\phi,\lambda) & = \zeta_0 \cos\phi \exp[-(\phi-\phi_m)^2/\delta^2)]\cos(k_i\lambda).
\end{align}
\end{subequations}
To mimic Earth's upper troposphere, we choose
\begin{equation}\label{eq:earth_param}
A = 3.4\times10^{-1}, B=4.1\times10^{-1}, C=4.0 \mbox{, and } D=2.3\times10^{-1}
\end{equation}
The corresponding dimensionalized flow on a sphere of Earth's radius and rotation rate resembles the midlatitude jet in the upper troposphere. It has a maximum westward velocity of ${\sim}33 \unit{m\,s^{-1}}$ at a latitude of ${\sim}40\degree$ and a maximum eastward velocity of ${\sim}22 \unit{m\,s^{-1}}$ at the equator; its zero is near ${\sim}17 \degree$. This zonal flow is barotropically stable. The eddy vorticity $\zeta_i$ decays meridionally away from its maximum absolute value $\zeta_0 \cos\phi_m$ at latitude $\phi_m = \pi/4 = 45\degree$, with characteristic meridional decay scale $\delta = \pi/9$. Its zonal wavenumber $k_i = 6$ is close to the dominant zonal wavenumber of transient eddies on Earth \citep[e.g.,][]{boer1983,straus1999}, which approximately coincides with the baroclinically most unstable zonal wavenumber  \citep[e.g.,][]{simmons1976,simmons1978,thorncroft1993,tim2009}.

We let the flow evolve freely without forcing or dissipation, apart from hyperviscosity, and analyze the time evolution of the mean flow and the eddies. We compare CE2 to the statistics of fully nonlinear simulations for different choices of parameters. To identify relevant nondimensional parameters controlling the evolution of the flow and the adequacy of CE2 closures, we rescale the mean and eddy vorticity equations (\ref{eq:vort_eddy_mean}), using the relative vorticity $2\mathrm{Ro}$ 
of the initial zonal flow $U$. Dimensionally, we have $\mathrm{Ro} = \Lambda/(2\Omega)$, where $\Lambda \approx 2\,\max(U)/a$ is the typical initial zonal-mean flow vorticity. Hence, $\mathrm{Ro}$ is a Rossby number measuring the mean flow vorticity to the equatorial planetary vorticity $2\Omega$. We measure the initial maximum vorticity $\zeta_0$ of the eddies relative to the mean-flow vorticity $2\mathrm{Ro}$ through the amplitude parameter $\epsilon = \zeta_0 \cos\phi_m / 2\mathrm{Ro}$. The quantities $\overline{\zeta}$, $\zeta'$, $\overline{\psi}$, $\psi'$, and $t$ are then rescaled with $4\pi\mathrm{Ro}$, $4\pi\epsilon\mathrm{Ro}$, $4\pi\mathrm{Ro}$, $4\pi\epsilon \mathrm{Ro}$, and $(4\pi\mathrm{Ro})^{-1}$, respectively. The equations of motion for the mean-flow and the eddies become
\begin{subequations}\label{eq:stream_adim}
    \begin{align}
    \frac{\partial\overline{\zeta}}{\partial t} &= -\epsilon^2 \overline{J_{n}\left(\psi',\zeta'\right)}, \\
    \mathrm{Ro}\left(\frac{\partial}{\partial t} - \frac{1}{\cos\phi}\frac{\partial \overline{\psi}}{\partial \phi}\frac{\partial}{\partial \lambda} \right) \zeta' &= 
    -\epsilon \mathrm{Ro} \left[J_{n}\left( \psi',\zeta'\right) - \overline{J_{n}\left( \psi',\zeta'\right)} \right]
    - \frac{\partial{\psi'}}{\partial \lambda} \left(1 + \mathrm{Ro}\frac{1}{\cos\phi}\frac{\partial\overline{\zeta}}{\partial \phi}\right),
    \end{align}
\end{subequations}
Equation \eqref{eq:stream_adim} gives the relative amplitude of the different terms if we assume that the typical length scales of the mean flow and eddies are of the order of the radius of the planet. An immediate simplification that results for small Rossby number $\mathrm{Ro}$ (the case we will consider) is that the advection of mean-flow vorticity $\bar\zeta$ by meridional velocity fluctuations is negligible compared with the $\beta$-term, which is a factor $\mathrm{Ro}^{-1}$ larger. 

The nondimensional parameters $\mathrm{Ro}$ and $\epsilon$ control the vorticity of the mean flow and of the eddies and are important for the evolution of the barotropic flow. Eddy--mean flow interactions are of order $\mathrm{Ro}$, and eddy-eddy interactions of order $\epsilon \mathrm{Ro}$, provided $\nabla_n^2$ is of order one. This is the case initially; however, it does not remain true over the evolution of the flow, as small scales are generated. The two parameters $\mathrm{Ro}$ and $\epsilon$ can be varied independently in our setup. In what follows, we explore how these parameters affect the flow evolution and the adequacy of CE2 and QL approaches in capturing it.

\subsection{Results}

        \subsubsection*{Varying eddy amplitudes}
        
For a fixed initial mean-flow Rossby number $\mathrm{Ro} \approx 0.06$ (corresponding to the Earth-like parameters in equation \eqref{eq:earth_param}), we run eddy lifecycle experiments for larger-amplitude initial eddies with $\epsilon \approx 6$, and for smaller-amplitude initial eddies with $\epsilon \approx 2$. The expectation is that CE2 and QL approaches are more successful for the smaller-amplitude eddies, for which the nonlinear eddy--eddy interactions (of order $\epsilon \mathrm{Ro}$) are weaker, and this is indeed borne out in the simulations. It is instructive to see in what way they fail to capture aspects of the flow evolution for the larger-amplitude eddies.

\begin{figure}[ht]
  \includegraphics[width=\textwidth]{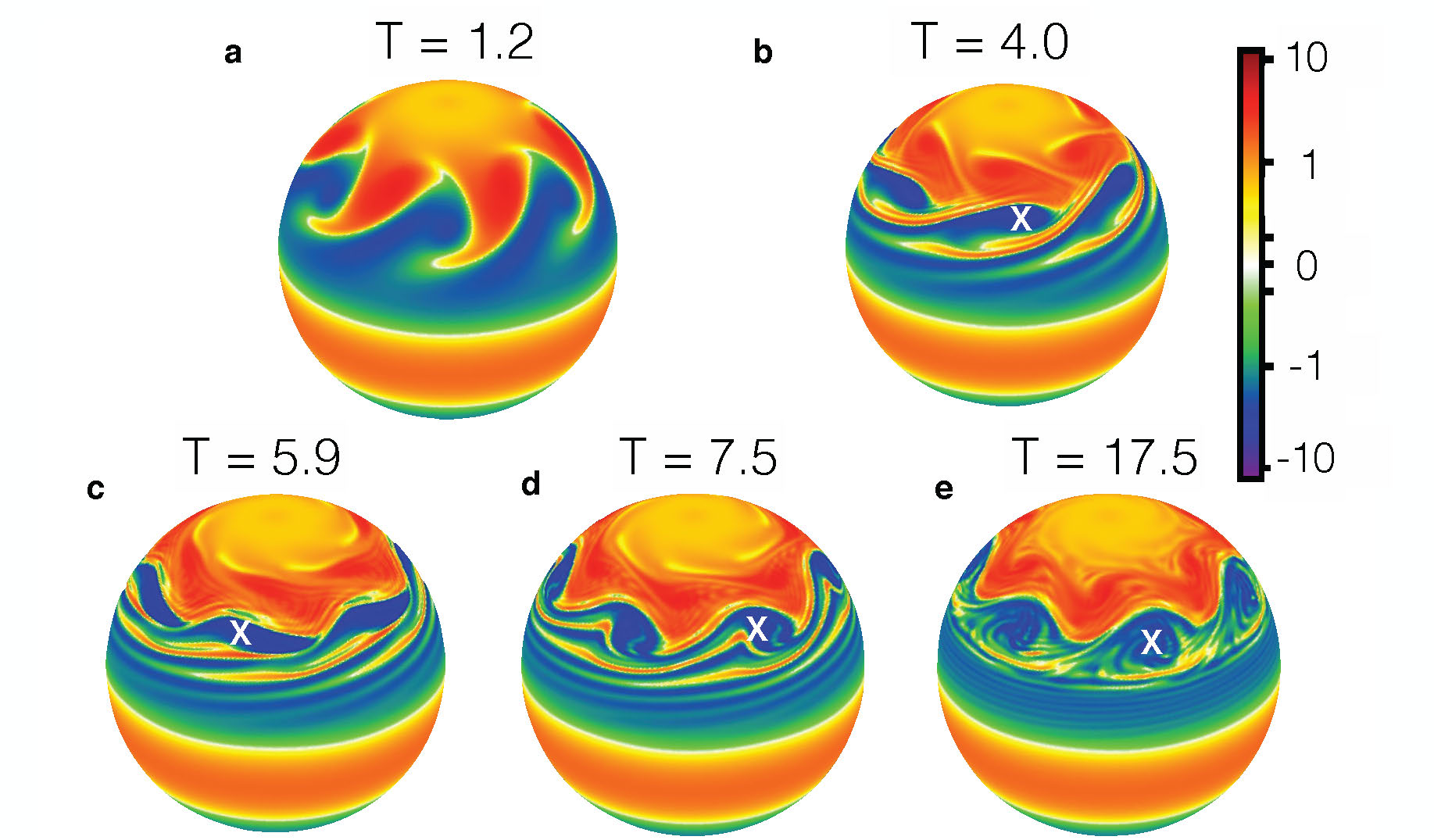}\\
  \caption{Evolution of relative vorticity in the fully nonlinear simulation for the larger-amplitude eddies ($\epsilon = 6$) for an Earth-like setting ($\mathrm{Ro} = 0.06$). The relative vorticity maps show the formation of cats' eyes,  with vorticity filaments rolling up within them. White X's mark the centres of some cats' eyes. Time scales are non-dimensionalized with the length of the day $2\pi \Omega^{-1}$. }\label{fig:EarthLike}
\end{figure}

For the larger-amplitude eddies, Fig.~\ref{fig:EarthLike} shows the relative vorticity $\zeta$ at 5 instants during the evolution of the flow. It is evident that the initial disturbance quickly becomes nonlinear and develops drawn-out filaments on the equatorward flank of the zonal jet. The filaments roll-up anticyclonically within cats' eyes structures (marked by X's in Fig.~\ref{fig:EarthLike}) that continue to have the initial zonal wavenumber $k_i = 6$. Such cats' eyes are characteristic of Rossby waves that break in ``surf zones'' around their critical layers\footnote{Fig.~\ref{fig:EarthLike}e shows some spurious oscillations in the critical layer region. This is the consequence of a too weak hyperdiffusion. With  hyperdiffusion strong enough to remove the ripples, we observed have noticable hyperdiffusive wave absorption over the time scales considered. This obscures the wave absorption due to eddy--eddy interactions and blurs the differences between fully nonlinear and CE2 simulations. Removing the spurious ripples while showing sharp differences between fully nonlinear and CE2 simulations would have required a much higher resolution, or perhaps a different numerical advection scheme.} \citep{stewartson1977,warn1978,McIntyre83,killworth1985,held1987}. 

\begin{figure}[ht]  
\noindent\includegraphics[width=\textwidth]{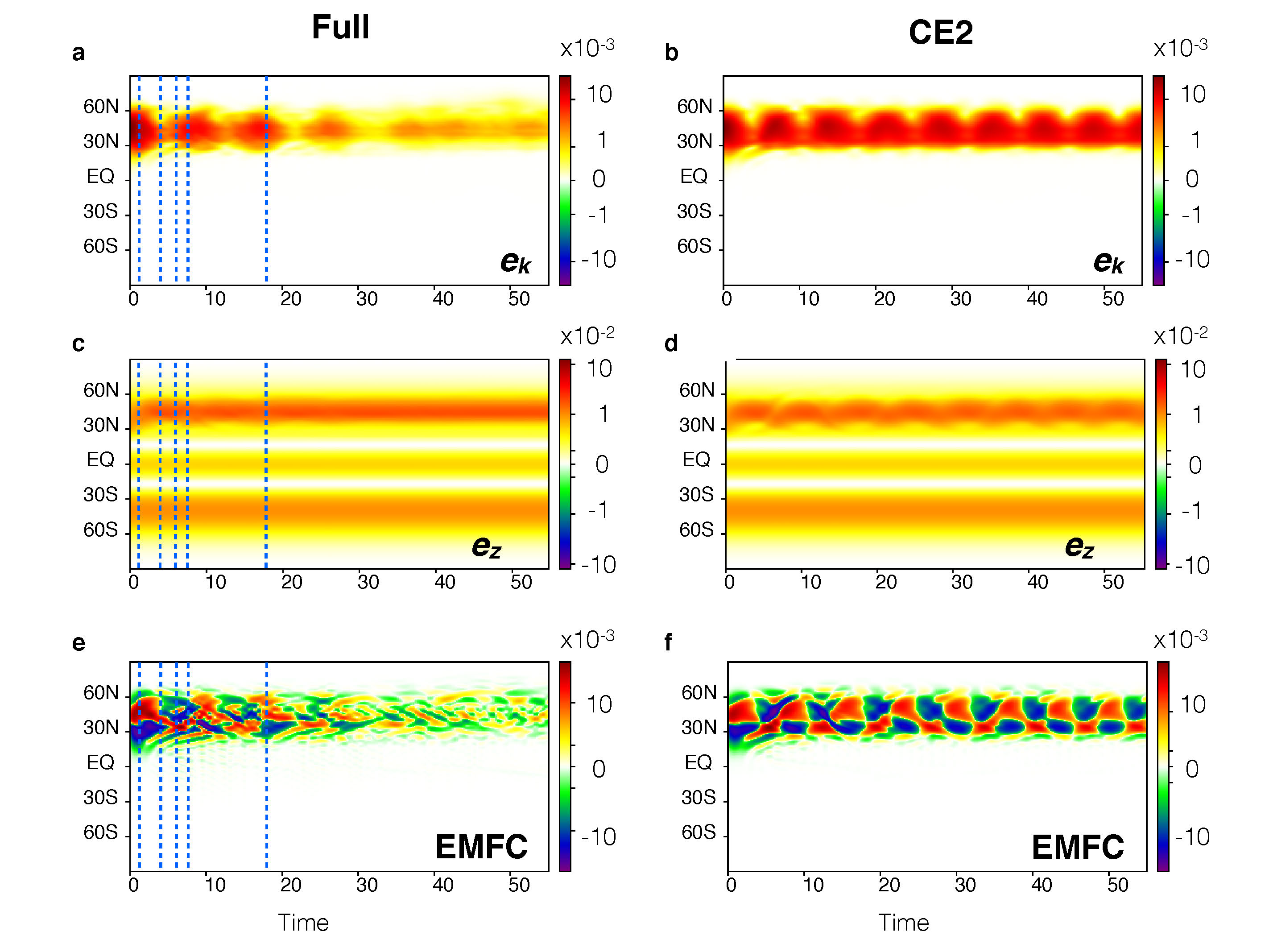}\\
\caption{Evolution of kinetic energy and eddy momentum flux convergence (EMFC) in the fully nonlinear simulation (left column) and in a direct CE2 calculation of the statistics (right column) for an Earth-like setting ($\mathrm{Ro} = 0.06$) and for the larger-amplitude eddies ($\epsilon = 6$). (a, d) Eddy kinetic energy $e_K$. (b, e) Zonal kinetic energy $e_Z$. (c, f) EMFC. Vertical blue dashed lines indicate at which times relative vorticity snapshots are shown in Fig.~\ref{fig:EarthLike}. Time scales and length scales are non-dimensionalized with the length of the day $2\pi \Omega^{-1}$ and the planet radius $a$. }\label{fig:fullvsCE2_strong}
\end{figure}

As the eddies break and eventually dissipate in the surf zone, they are absorbed by the mean flow, and their kinetic energy decays. The total eddy kinetic $E_K = 0.5\int_{\phi} e_K \cos\phi\,d\phi$, where $e_K = 0.5 (\overline{u'^2} + \overline{v'^2})$, becomes very small at large times ($\gtrsim 30$, see Fig.~\ref{fig:fullvsCE2_strong}a). At those times, most of the initial kinetic energy has been transferred to the mean zonal flow. The local zonal kinetic energy $e_Z = 0.5 \overline{u}^2$ ($\bar u \equiv -\partial_\phi \bar\psi$ of the mean zonal flow increases in the core of the midlatitude jet, roughly in proportion to the decrease of the eddy kinetic energy $e_K$ (Fig.~\ref{fig:fullvsCE2_strong}c). Total energy $E_K + E_Z$, with $E_Z = 0.5\int_{\phi} e_Z \cos\phi\,d\phi$, is approximately conserved in the model, up to very small losses (${\sim} 0.2\%$ of the total after a time of 50) through subgrid-scale dissipation. That is, although dissipation at small scales in the surf zone is essential to generate irreversibility, the kinetic energy loss associated with it is small compared to the transfer to the mean flow. 

The transfer of $E_K$ to $E_Z$ implies acceleration of the mean zonal jet. This acceleration occurs through transfer of momentum from the eddies to the mean flow, as can be seen from the nondimensionalized zonally averaged momentum equation \eqref{eq:barotropic_momentum} in the inviscid limit ($\mathcal{F}_{\mathbf{v}}=0$):
\begin{equation}\label{eq:barotropic_zonal_mom}
    \cos\phi \frac{\partial \bar{u}}{\partial t} = - \frac{1}{\cos\phi}\frac{\partial}{\partial \phi}\left[\cos^2\phi\, \overline{u'v'} \right]. 
\end{equation}
Acceleration of the mean zonal flow occurs where eddy momentum fluxes $\overline{u'v'}\cos\phi$ converge. Multiplying the mean zonal momentum equation (\ref{eq:barotropic_zonal_mom}) by $\bar u$ and integrating by parts yields the equation for the zonal kinetic energy $E_Z$,
\begin{equation}\label{eq:barotropic_energy}
    \frac{d}{dt}E_Z = \int_{\phi} \cos^2\phi\, \overline{u'v'}\frac{\partial}{\partial{\phi}} \left(\frac{\bar{u}}{\cos\phi}\right)\,d\phi =-\frac{d}{dt}E_K\mbox{,}
\end{equation}
where the right-hand side is obtained from a corresponding integral of the eddy momentum equations. This shows that transfer of kinetic energy from the eddies to the mean flow occurs through eddy momentum fluxes that are up the gradient of angular velocity $\bar{u} \cos^{-1}\phi$. The acceleration of the mean zonal jet at its core (Fig.~\ref{fig:fullvsCE2_strong}c) thus is associated with eddy momentum flux convergence $\partial_\phi(\overline{u'v'}\cos\phi)$ (EMFC, see Fig.~\ref{fig:fullvsCE2_strong}e). 

However, the eddy kinetic energy does not decay monotonically. Instead, it exhibits damped oscillations during which eddy momentum fluxes cause zonal angular momentum to slosh back and forth meridionally within the jet (Fig.~\ref{fig:fullvsCE2_strong}e). The alternating poleward and equatorward momentum fluxes (with decreasing amplitude) on the equatorward flank of the jet are result of nonlinear processes within the surf zone. These processes  have been described in an idealized analytical model of Rossby-wave breaking in critical layers, the Stewartson-Warn-Warn (SWW) theory (\citealp{stewartson1977,warn1978}; see also \citealp{killworth1985}). The oscillation on the poleward flank of the jet may be more linear, originating from the reflection of Rossby wavepackets from reflecting levels that arise because $\beta$ decreases with latitude \citep[e.g.,][]{lorenz2014}. 

Figure~\ref{fig:fullvsCE2_strong} (right column) shows the kinetic energies and EMFC obtained from a direct calculation of these statistics with CE2. CE2 captures the oscillation of kinetic energy between eddies and the mean zonal flow, with a period similar to the fully nonlinear simulation (Fig.~\ref{fig:fullvsCE2_strong}a, b). However, the oscillations are too weakly damped; large eddy kinetic energies $e_K$ persist for a long time. The eddy absorption in the surf zone is not adequately captured by CE2 because CE2 cannot resolve the generation of small scales in the surf zone through eddy--eddy interactions. Consistently, unrealistically strong oscillations persist in the EMFC under CE2 (Fig.~\ref{fig:fullvsCE2_strong}f). How these oscillations arise from the perspective of wave mechanics, and why CE2 cannot capture the wave absorption in this case, is illustrated in \ref{a:QL-large} in a QL simulation that corresponds to the CE2 calculations shown here. The phenomenology of such oscillations has been described analytically by \cite{haynes1987} in the context of a QL truncation of the SWW model.

\begin{figure}[ht]
  \noindent\includegraphics[width=\textwidth]{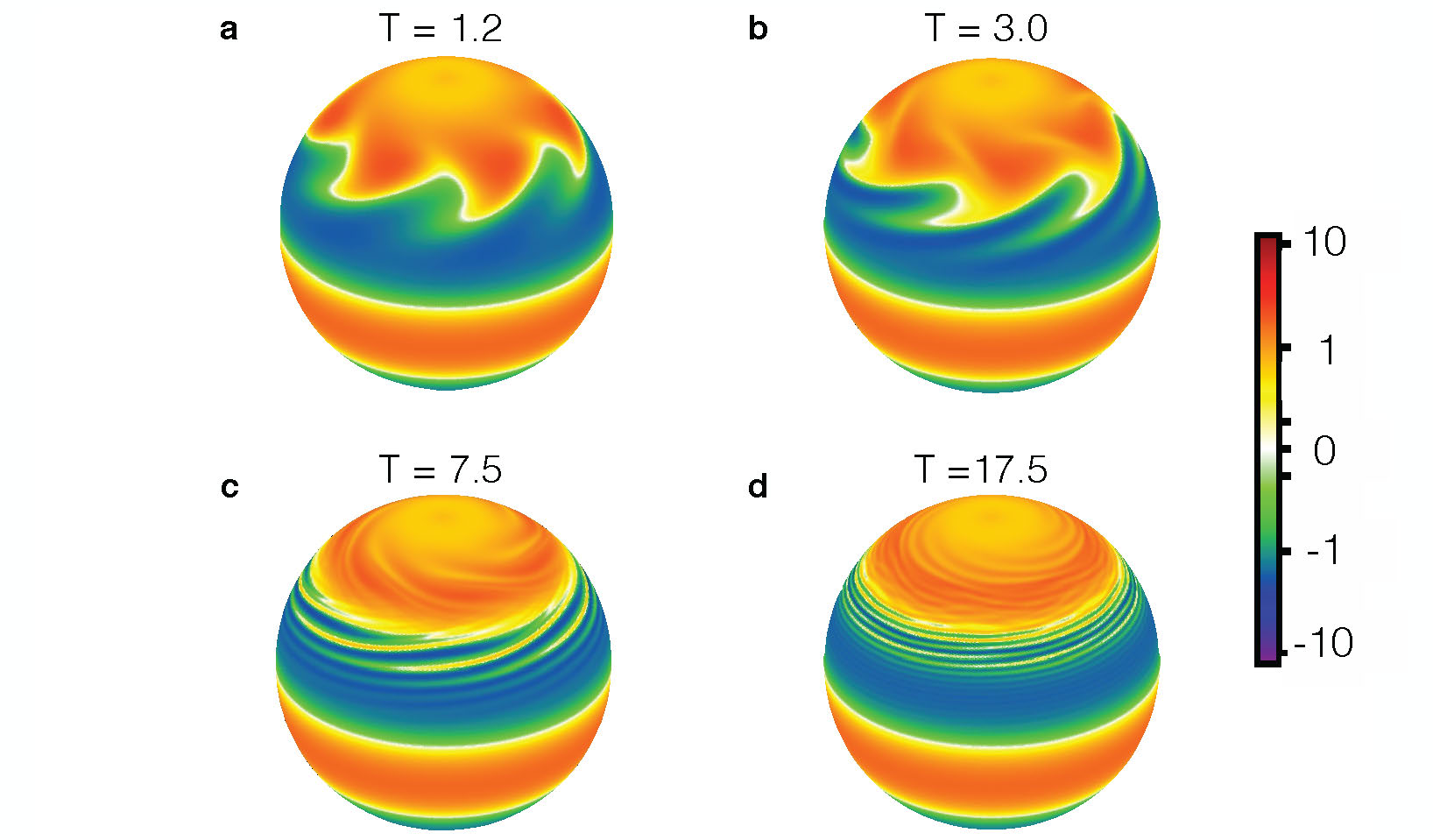}\\
  \caption{Evolution of relative vorticity in the fully nonlinear simulation for the smaller-amplitude eddies ($\epsilon = 2$) for an Earth-like setting ($\mathrm{Ro} = 0.06$). Time scales are non-dimensionalized with the length of the day $2\pi \Omega^{-1}$. }\label{fig:EarthLike_low}
\end{figure}

\begin{figure}[ht]
  \noindent\includegraphics[width=\textwidth]{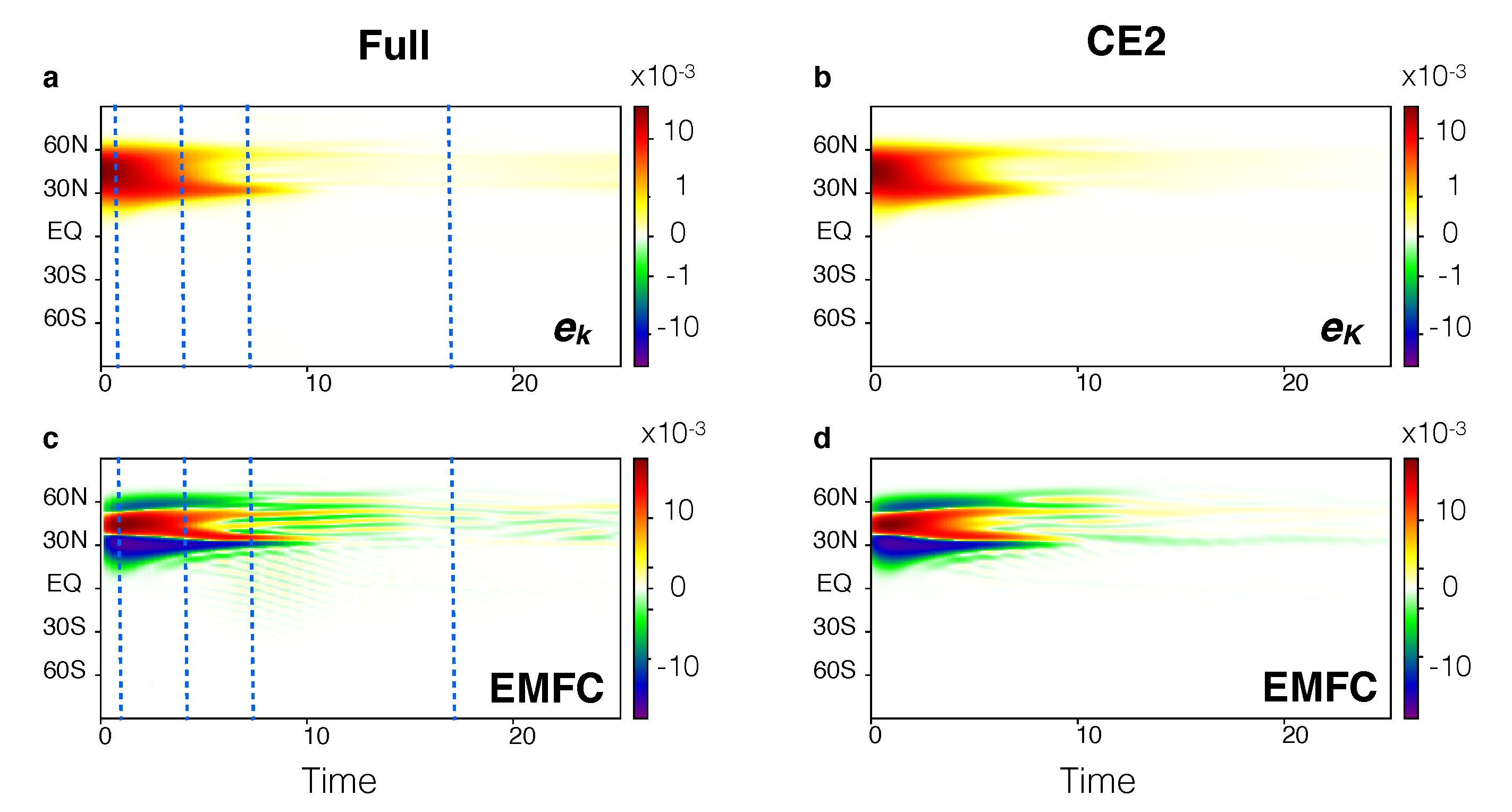}\\
  \caption{Evolution of kinetic energy and eddy momentum flux convergence (EMFC) in the fully nonlinear simulation (left column) and in a direct CE2 calculation of the statistics (right column) for an Earth-like setting ($\mathrm{Ro} = 0.06$) and for the larger-amplitude eddies ($\epsilon = 2$). (a, d) Eddy kinetic energy $e_K$. (c, f) EMFC. Vertical blue dashed lines indicate the times corresponding to relative vorticity snapshots on Fig.~\ref{fig:EarthLike}. Time scales are non-dimensionalized with the length of the day $2\pi \Omega^{-1}$ and length scales with the planet radius $a$.}\label{fig:fullvsCE2_weak}
\end{figure}
       
Eddy--eddy interactions are weaker and CE2 is more successful in capturing the flow dynamics when the amplitude of the initial perturbation is decreased by a factor 3 ($\epsilon \approx 2$). Cats' eyes with rolling-up vorticity filaments do not develop in the fully nonlinear simulation (Fig.~\ref{fig:EarthLike_low}). Instead, eddies are sheared by the mean flow, which transfers eddy kinetic energy $e_K$ to the mean flow through the Orr mechanism, which is only weakly nonlinear because it involves the interaction of disturbances with the slowly varying mean flow \citep[e.g.][]{thomson1887,orr1907,farrell1987,bouchet2013}. The transfer of eddy kinetic energy to the mean flow occurs over time scales of a couple days, corresponding to the shear time scale of the mean zonal flow (Fig.~\ref{fig:fullvsCE2_weak}). The damped oscillatory behavior seen in the larger-amplitude simulation disappears. Because eddy absorption results from the mean flow shearing the eddies---an eddy--mean flow interaction that is captured by CE2---statistics calculated directly with CE2 come in very close agreement with those from the fully nonlinear simulation (Fig.~\ref{fig:fullvsCE2_weak}, right column). As in the nonlinear case, eddy absorption occurs through the formation of small-scale vorticity filaments. But instead of rolling up inside cat's eyes, here they stretch around the planet.

It is worth noting that eddies are also sheared equatorward of the cats' eyes in the larger-amplitude simulation (Fig.~\ref{fig:EarthLike}). Hence, weakly nonlinear eddy absorption also occurs in this simulation, but it is not the dominant effect responsible for eddy absorption.

\paragraph*{Relative amplitude of terms in the potential vorticity budget}

For the larger-amplitude experiment, the evident importance of the development of small scales through eddy--eddy interactions seems consistent with the order of magnitude of the terms in the vorticity equations \eqref{eq:stream_adim}. The eddy--eddy interactions appear of order $\epsilon \mathrm{Ro} \approx 0.4$, compared with the $\beta$-term which is responsible for  Rossby wave dynamics and is of order 1. Hence, the eddy--eddy interactions are not negligible compared with Rossby wave dynamics. By contrast, the interactions of the disturbance with the mean flow shear are of order $\mathrm{Ro} \approx 0.06$ and hence are much weaker.

For the smaller-amplitude experiment, the dimensional analysis of the vorticity equations \eqref{eq:stream_adim} suggests that the eddy--eddy interactions now are of order $\epsilon \mathrm{Ro}\lesssim 0.2$ compared with the $\beta$-term. Hence, the eddy--eddy interactions become close to being negligible compared with Rossby wave dynamics, consistent with the simulation results. However, the dimensional analysis also suggests that interactions of the disturbance with the mean flow shear still are of order $\mathrm{Ro} \approx 0.06$ and hence are weaker still, albeit of the same order of magnitude as the eddy--eddy interactions. Yet the eddy--eddy interactions are inhibited in the fully nonlinear simulation, whereas the shear interactions dominate the eddy absorption, illustrating the limits of dimensional analysis in this nonlinear problem.

Equation \eqref{eq:stream_adim} is useful to determine which parameter to vary. Nevertheless, one has to be careful when interpreting the relative amplitude of the different terms. A small term can be fundamental for the dynamics. For example, the SWW theory has shown that eddy--eddy interactions, albeit weak, can determine at leading order momentum fluxes because they dominate the vorticity budget in a thin critical layer, where linear theory breaks down. This remains true in the limit where the relative amplitude of the eddy-eddy interactions goes to zero. Moreover, the relative amplitude of the terms in the vorticity budget evolves with time as small scales develop. 
      
\subsubsection*{Varying Rossby numbers}
        
To illustrate how variations of the mean-flow Rossby number affect the evolution of disturbances, we use larger-amplitude eddies ($\epsilon \approx 6$) and decrease the Rossby number $\mathrm{Ro}$ from $0.06$ to $0.02$. Dimensionally, this is equivalent to weakening the initial flow while keeping the rotation rate of the planet constant, or to increasing the rotation rate while keeping the initial flow constant. Based on the dimensional analysis of the vorticity equations \eqref{eq:stream_adim}, this reduction of the Rossby number should decrease the relative magnitude both of eddy--eddy interactions and of shearing of eddies by the mean flow relative to the $\beta$-term, maintaining the relative amplitude of the eddy--eddy interactions to the $\beta$-term.

\begin{figure}[ht]
  \noindent\includegraphics[width=\textwidth]{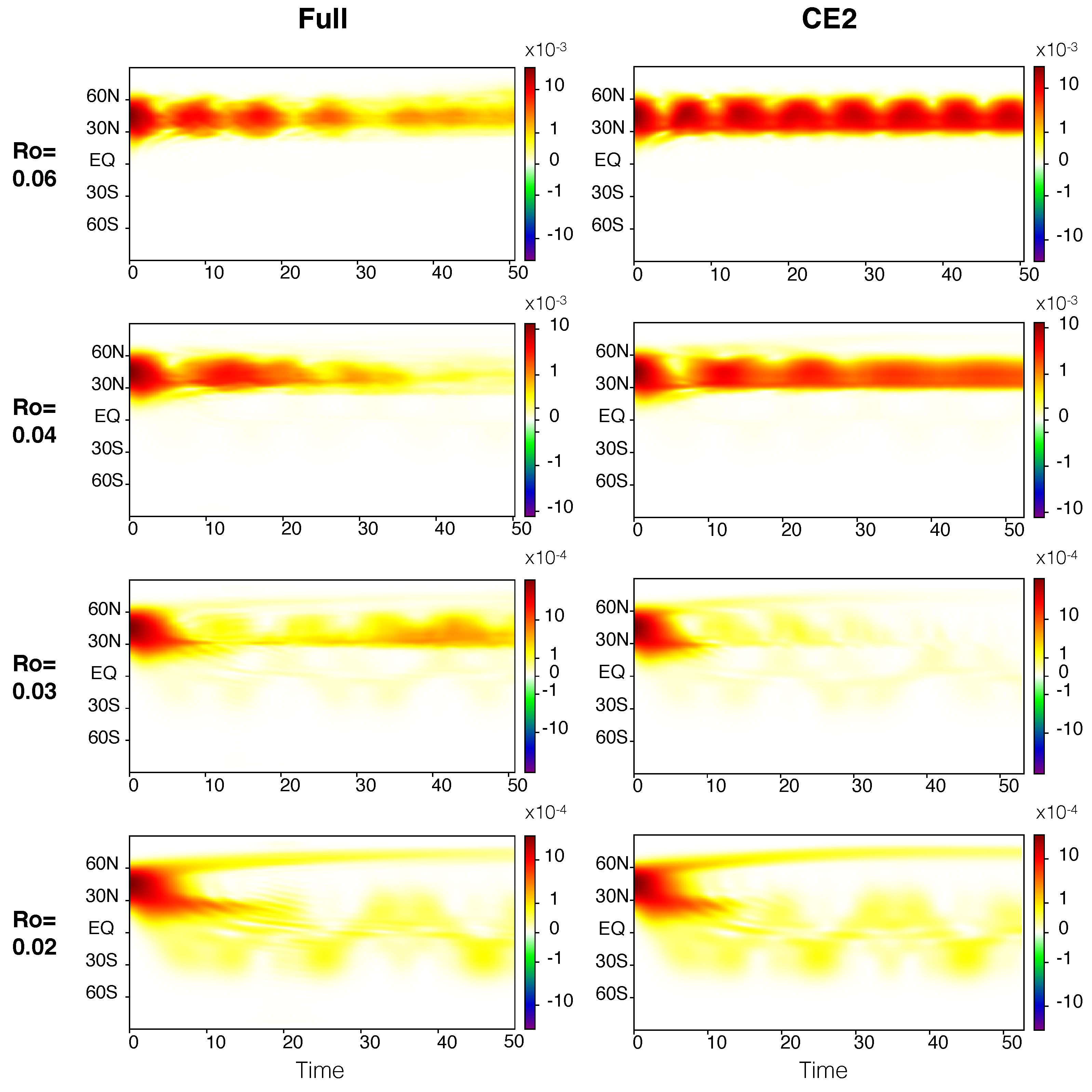}\\
  \caption{Evolution of eddy kinetic energies $e_K$ in the fully nonlinear simulation (left column) and in a direct CE2 calculation of $e_K$ (right column) for the larger-amplitude eddies ($\epsilon = 6$), with Rossby number decreasing from $\mathrm{Ro} = 0.06$ to $0.02$. Time scales are non-dimensionalized with the length of the day $2\pi \Omega^{-1}$ and length scales with the planet radius $a$.}\label{fig:fullvsCE2_rot_eke}
\end{figure}

\begin{figure}[ht]
  \noindent\includegraphics[width=\textwidth]{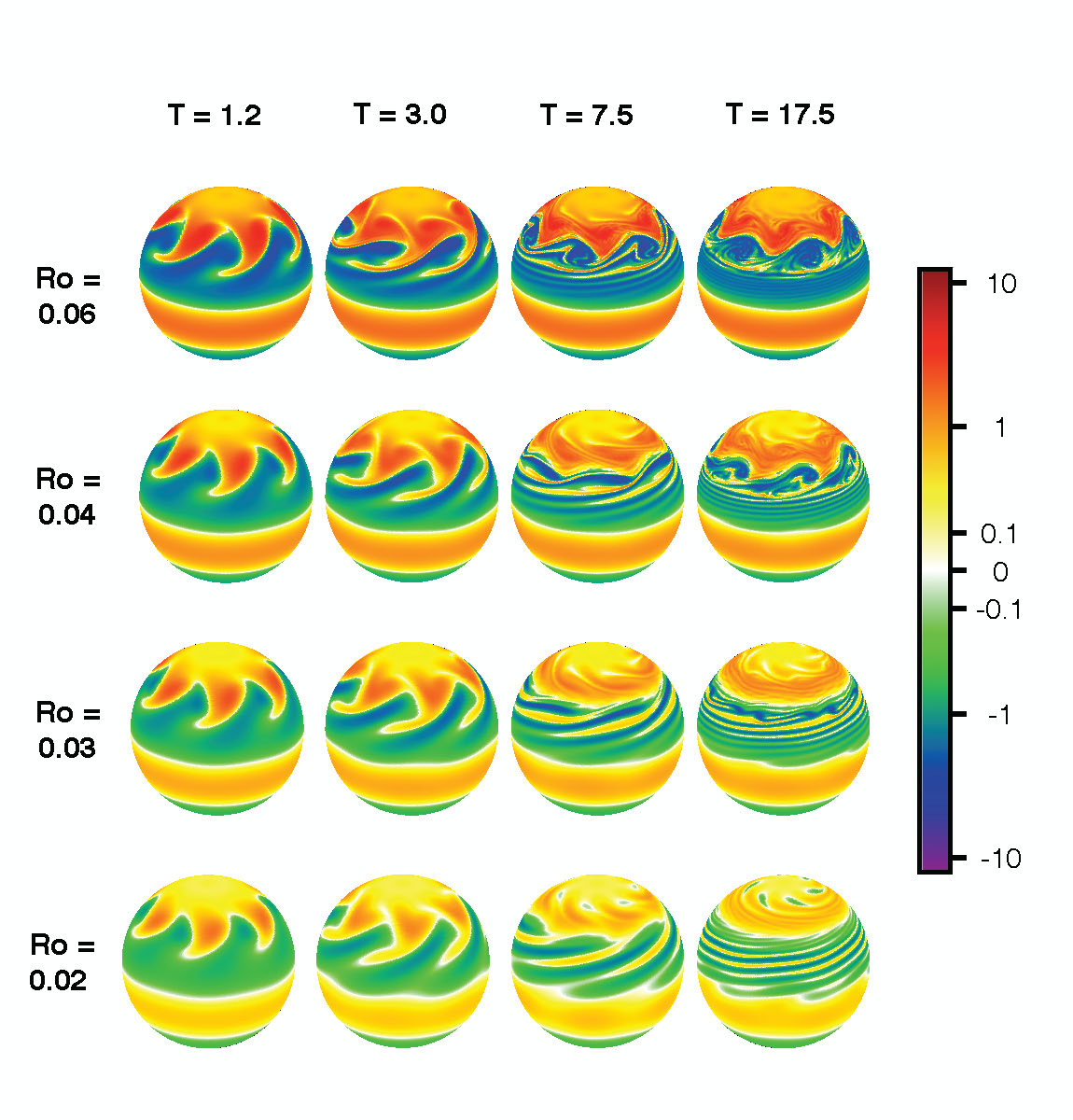}\\
  \caption{Evolution of relative vorticity in the fully nonlinear simulation for the larger-amplitude eddies ($\epsilon = 6$), with Rossby number decreasing from $\mathrm{Ro} = 0.06$ to $0.02$. For comparison, the first row reproduces the vorticity maps already shown in Fig.~\ref{fig:EarthLike}. Time scales are non-dimensionalized with the length of the day $2\pi \Omega^{-1}$.}
\label{fig:fullvsCE2_rot}
\end{figure}

Figure~\ref{fig:fullvsCE2_rot_eke} compares the time evolution of the eddy kinetic energy $e_K$ in the fully nonlinear and the CE2 simulations. As we have already seen, eddy absorption at $\mathrm{Ro} = 0.06$ is not captured by CE2 because it is strongly nonlinear. However, at the smaller Rossby number $\mathrm{Ro} = 0.02$, it is faithfully captured by CE2. Indeed, eddy absorption appears more weakly nonlinear for $\mathrm{Ro} = 0.02$, and dominated by mean-flow shearing,  as is evident on the relative vorticity maps (Fig.~\ref{fig:fullvsCE2_rot}). Vorticity maps for $\mathrm{Ro} = 0.04$ and $\mathrm{Ro} = 0.03$ show the transition from weakly to strongly nonlinear absorption: the meridional extent of the nonlinear surf zone decreases with increasing rotation, while shearing effects become more important (Fig.~\ref{fig:fullvsCE2_rot}). Because CE2 can capture the weakly nonlinear shearing interactions but not the strongly nonlinear eddy--eddy interactions in the surf zone, it becomes gradually more adequate as the rotation rate increases (Fig.~\ref{fig:fullvsCE2_rot_eke}).  This occurs although the orders of magnitude of the relevant terms suggested by the dimensional analysis decrease by equal factors of order $\mathrm{Ro}$ as the mean-flow Rossby number decreases. It appears that what is important here is that the magnitude of the advection of absolute vorticity, and hence of linear Rossby wave dynamics, increases relative to both of these terms. For $\mathrm{Ro = 0.02}$, shear explains eddy decay and jet acceleration, even though the nondimensionalization \eqref{eq:stream_adim} suggests shear should be much smaller than eddy-eddy interactions. 

   \subsubsection*{Higher-order closures}

CE2 fails to capture eddy absorption when eddy--eddy interactions are important for the dynamics. We tested whether a higher-order closure (CE3*) captures eddy absorption more faithfully. CE3* is described in \cite{marston2014}. It truncates the cumulant equations at third order, ensuring realizability by projecting out modes with (unphysical) negative energies.

\begin{figure}[ht]
  \noindent\includegraphics[width=\textwidth]{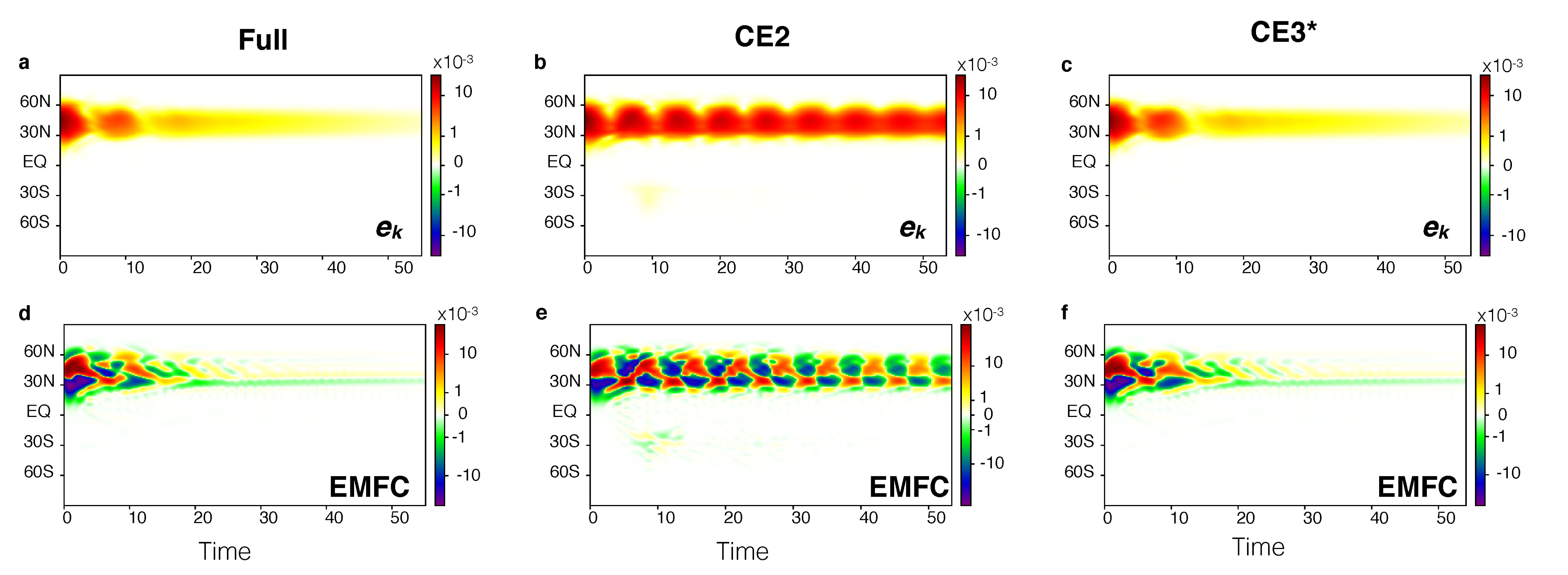}\\
  \caption{Evolution of eddy kinetic energies $e_K$ (top) and eddy momentum fluxes EMFC (bottom) in the fully nonlinear simulation (left column), in a direct CE2 calculation (middle), and in a CE3* calculation (right), all for the larger-amplitude eddies ($\epsilon = 6$) and an Earth-like setting ($\mathrm{Ro} = 0.06$). Compared with the simulations in previous figures, the resolution is reduced to $M=40$ and $L=20$.}
\label{fig:CE3}
\end{figure}

The results are summarized in Fig.~\ref{fig:CE3}. Because CE3* is computationally expensive, the resolution has been reduced to $M=40$ and $L=20$. For simplicity we turn off eddy damping ($\tau = \infty$, see \cite{marston2014} for definitions). The full and CE2 simulations in Fig.~\ref{fig:CE3} are run at this lower resolution and are consistent with the higher-resolution runs (Fig.~\ref{fig:fullvsCE2_strong}); however, they do exhibit a faster eddy damping because of the stronger diffusion. It appears that CE3* captures the eddy absorption very accurately. We also tested other closures that take into account third-order terms \citep{marston2014}; they bring similar improvements. 

\subsection{Implications}

The results show that direct CE2 calculation of barotropic flow statistics representative of the upper troposphere can succeed in circumstances when the dominant nonlinear interaction is between eddies and the mean flow, for example, by shearing. They fail when strongly nonlinear eddy--eddy interactions become important in surf zones around critical layers, where the roll-up of vorticity filaments leads to the generation of small scales. This is a process that cannot be captured in CE2. However, higher-order closures, which take some effects of third cumulants on second cumulants into account, can perform better in such cases---at the price of increased conceptual and computational complexity. 

Weakly nonlinear eddy--mean flow interactions seem to be favored over strongly nonlinear eddy--eddy interactions when the eddy vorticity is small enough compared with the planetary vorticity. The transition from weakly to strongly nonlinear interactions predominating occurs above a critical value of eddy amplitude $\epsilon$ that is a decreasing function of the Rossby number $\mathrm{Ro}$. The parameter $\epsilon$ need not be small for absorption through weakly nonlinear shearing to occur. For example, for small $\mathrm{Ro}$, weakly nonlinear eddy absorption through shearing seems to be favored even when a large value of $\epsilon$ suggests that nonlinear eddy--eddy interactions should be larger than the mean-flow shearing of eddies. 

When strongly nonlinear eddy--eddy interactions are favored, high eddy kinetic energies develop in the QL/CE2 approximation because momentum sloshes back and forth meridionally within the jet, without sufficient absorption. Lifecycle experiments carried out with a QL GCM have shown that this phenomenology is relevant to the upper troposphere in an Earth-like setting, highlighting the relevance of the simplified barotropic model: Earth's upper troposphere appears to be in a regime in which nonlinear eddy--eddy interactions in surf zones are important for the structure of the momentum fluxes \citep{ait2014}.

\section{Conclusions}

Atmospheric flows are highly anisotropic and inhomogeneous, with rich spatial structures.  Turbulent closures that respect the anisotropy and inhomogeneity may enable the direct statistical simulation of Earth's atmosphere \citep{marston2011}.  Expansion of statistics in equal-time cumulants yields equations of motion for the statistics that can already provide useful closures at second order (CE2), because the mean flows and interactions of perturbations with them are strong \citep{herring1963, ogorman2007, marston2008,tobias2013,marston2014}.  CE2 solves the first two cumulant (central moment) equations of the QL approximation, in which interactions between mean flows and fluctuations around it are retained, while nonlinear eddy--eddy interactions are neglected. In section~\ref{sec:CE2}, we formulated CE2 for the Boussinesq equations by introducing a condensed tensorial notation. The case of the anelastic equations is presented in \ref{a:anelastic} and involves the use of density weighted averages. We tested the relevance of CE2 to two distinct atmospheric flows involving different length and time scales and force balances: turbulent convection in the atmospheric boundary layer (section~\ref{sec:boundary_layer}), and weak two-dimensional turbulence representative of the upper troposphere (section~\ref{sec:barotropic}).  

Convection in the atmospheric boundary layer links large-scale atmospheric dynamics aloft to the surface underneath, mediating the exchanges of momentum, energy, and water between the surface and the free troposphere. Motions in boundary layers and in clouds that have their roots in them have dynamical scales of meters, meaning that they need to be parameterized in GCMs. Current parameterization schemes have numerous shortcomings; our inability to represent cloud and boundary layer dynamics adequately in climate models is the largest source of uncertainty in climate change projections. Because cumulant expansions capture interactions between fluctuations (e.g., thermals) and mean fields and take non-local correlations of fluctuations into account, without requiring the introduction of tunable parameters that proliferate in current parameterization schemes, they may offer a way to achieve more physically consistent parameterizations. We presented encouraging initial results, showing that a QL large-eddy simulation of a dry convective boundary layer captures the first-order statistics (e.g., mean boundary layer growth) of a corresponding fully nonlinear LES. However, it does not capture second-order statistics adequately. More work is required to investigate to what degree these results hold generally, in broader classes of boundary layer flows and in the presence of moisture effects and clouds, and how the QL results can be improved by including representations of higher-order effects, such as the turbulent transport of kinetic energy.

The potential for development of parameterisation schemes based on CE2 is a promising direction for future research. But it requires overcoming both technical and theoretical challenges. On the technical level, fast numerical methods are required for the method to be competitive with other subgrid schemes.  This could be achieved by dimensional reduction to capture only the most important non-local correlations in the second cumulant.  At the theoretical level, it is not clear to what extent CE2 and possible extensions will be able to describe moist convection, or what effects vertical shear will have on its accuracy.  It appears necessary to include some effects of eddy--eddy interactions, such as those captured by the third order CE3* approximation \citep{marston2014}. 

At the planetary scale, how and where eddies in the upper troposphere dissipate controls the strength and direction of momentum fluxes and thus climatic features such as surface winds. A one-layer barotropic model that mimics the behavior of the upper troposphere illustrates different mechanisms through which eddy absorption by the mean flow can occur. CE2 describes eddy absorption well when it occurs through shearing of eddies by the mean flow. This happens when the vorticity that characterizes the eddies is small compared with the planetary vorticity (planetary rotation rate). When the eddy vorticity is large, CE2 is not adequate because eddy absorption results from the formation of small scales that form through eddy--eddy interactions in critical layers. A comprehensive theory that describes these weakly and strongly nonlinear absorptive regimes is lacking.  

Our results suggest that, in general, higher-order closures are required for an accurate direct statistical simulation of large-scale and smaller-scale atmospheric flows. We have tested a few of them in the large-scale context and found improvement compared with CE2. Nevertheless, going beyond CE2 raises a number of questions. Higher-order closures are several orders of magnitude slower than CE2; currently, they are much slower than direct simulation of the flows, while CE2 can be faster that direct simulations. Dimensional reduction may be able to help here.   More generally, once eddy--eddy interactions are taken into account, the whole hierarchy of cumulants is active and not completely described by any finite closure.  Realizability of closures becomes an issue \citep[e.g.,][]{Orszag70a,Orszag73a, marston2014}, and it is known that intermittency cannot be adequately described in this way \citep{frisch1995,lesieur2008}.  But the existence of anisotropic and inhomogeneous mean flows provides a starting point for a systematic exploration of statistical closures.

\ack
We thank Kyle Pressel for helping in the development of the quasi-linear LES and Joe Skitka for sharing results about the oceanic boundary layer. We are grateful for helpful discussions with Freddy Bouchet, Greg Chini, Baylor Fox-Kemper, Cesare Nardini, and Steve Tobias. The comments of two anonymous reviewers improved the manuscript significantly. This work was supported by the U.S. National Science Foundation under grants CCF-1048575 (FAC and TS) and CCF-1048701 (FAC and JBM).

\appendix
\setcounter{section}{0}

\section{Cumulant expansion of anelastic flow}\label{a:anelastic}
 
Because the flow is non-divergent in the Boussinesq equations, it is possible to directly use any averaging operator satisfying the properties (\ref{eq:avg_pties}). However, for more general flows, in which the density may vary, the requirements on the averaging operator need to be modified so that second-order equations for fluctuations consistent with the conservation laws are obtained. Because density appears as a weight in all integrals of conserved quantities over the flow domain, the density weighted average $\overline{(\cdot)}^{*} = \overline{({\rho\cdot})}/{\overline{\rho}}$, with $\overline{(\cdot)}$ satisfying the Reynolds properties (\ref{eq:avg_pties}), leads to a decomposition of the flow that is amenable to closures such as CE2. The density weighted average satisfies (\ref{eq:avg_pties}a-c) but, importantly, not commutativity with partial differentiation (\ref{eq:avg_pties}d).  Thus, we can define an eddy mean flow decomposition 
\begin{subequations}\label{eq:favre}
    \begin{align}
       f & = \overline{f}^* + \hat{f}, \\
       \overline{fg}^* & = \overline{f}^*\overline{g}^* + \overline{\hat{f}\hat{g}}^*,    
    \end{align}
\end{subequations}
with hats $\hat{(\cdot)} = (\cdot) - \overline{(\cdot)}^*$ denoting fluctuations around the density weighted mean. This is sometimes called the Favre decomposition. 

How a varying density affects cumulants and CE closures can be illustrated with the anelastic equations, an extension of the Boussinesq equations that allows the reference density $\rho_0=\rho_0(z)$ to vary with height $z$. In the anelastic approximation, density perturbations $\delta \rho$, the pressure potential $\Phi = \delta p/\rho_0$, and the buoyancy $b = -g \delta \rho/\rho_0$ are defined relative to this height-dependent and hydrostatically balanced reference density. The state vector $\mathbf{\Psi}$ is defined as in (\ref{eq:psi}), but the augmented state vector $\widetilde{\mathbf{\Psi}}$ is now taken to be
\begin{equation}\label{eq:psi_anelastic}
    \mathbf{\tilde{\Psi}} = 
    \left(
        u\,,
        v\,,
        w\,,
        b\,,
        \nabla\Phi
    \right)^T \mbox{.}
\end{equation}
We have to consider covariances involving $\nabla\Phi$ and not $\Phi$ because multiplying by density does not commute with derivatives. For simplicity, we assume that the linear operator $\mathcal{L}$ is homogeneous in the reference density and satisfies $\rho_0\mathcal{L}(\mathbf{\tilde{\Psi}}) = \mathcal{L}(\rho_0 \mathbf{\tilde{\Psi}})$. It implies that it cannot involve any spatial derivative. The anelastic equations in a reference frame rotating with angular velocity $\mathbf{\Omega}$ are \citep{vallis2006}:
\begin{subequations}\label{eq:anelastic2}
    \begin{align}
        \frac{\partial \rho_0 \mathbf{\Psi}}{\partial t} + 
        \nabla \cdot [\rho_0 \mathbf{\Psi} \otimes \mathbf{u}] &  
        = \mathbf{L}[ \rho_0 \mathbf{\Psi} ] - 
        \rho_0 \nabla \Phi + \mathbf{F}, \\
        \nabla \cdot \mathbf{\rho_0} \mathbf{u} & = 0.
    \end{align}
\end{subequations}
The Boussinesq equations (\ref{eq:Boussinesq}) are obtained from the anelastic equations by setting $\rho_0$ to a constant. The continuity equation (\ref{eq:anelastic2}b) again reduces to a non-divergence constraint. Because the reference density $\rho_0$ appears in it, we need to use the reference-density weighted average $\overline{(\cdot)}^{*} = \overline{({\rho_0\cdot})}/{\Bar{\rho}_0}$ to derive the cumulant expansion. 

The equation for the first cumulant $\overline{\mathbf{\Psi}(\mathbf{r},t)}^*$ is obtained by averaging (\ref{eq:anelastic2}):
\begin{subequations}\label{eq:anelastic_cum1_bis}
    \begin{align}
        \frac{\partial\Bar{\rho}_0\overline{\mathbf{\Psi}}^*}{\partial t} 
        + \nabla\cdot[\Bar{\rho}_0\overline{\mathbf{\Psi}}^* \otimes \overline{\mathbf{u}}^*] 
        = -\nabla \cdot[\Bar{\rho}_0\overline{\hat{\mathbf{\Psi}} \otimes \hat{\mathbf{u}}}^*]  
        + \mathbf{L}[\Bar{\rho}_0\mbox{ }\overline{\mathbf{\Psi}}^*]
        -\bar \rho_0 \overline{\nabla \phi}^*,
        \\
        \nabla \cdot \Bar{\rho}_0\mbox{ }\overline{\mathbf{u}}^* = 0. & &  
    \end{align}
\end{subequations}

To define a second central moment or second cumulant, we need a quantity that respects the symmetry (\ref{eq:symmetry}) of the covariance matrix under exchange of the spatial coordinates $\mathbf{r_1}$ and $\mathbf{r_2}$, that gives the second-order term in the first cumulant equation (\ref{eq:anelastic_cum1_bis}), and that corresponds to a density weighted average when $\mathbf{r_1} = \mathbf{r_2}$. A possible choice is 
\begin{equation}
\mathbf{C_{+}}(\mathbf{r_1},\mathbf{r_2},t) = \frac{1}{2}\overline{
\left[\rho_0(\mathbf{r_1}) + \rho_0(\mathbf{r_2})\right]
\hat{\mathbf{\Psi}}(\mathbf{r_1}, t) \otimes \hat{\mathbf{\Psi}}(\mathbf{r_2},t)}.
\end{equation}
We also define the two auxiliary covariances corresponding to (\ref{eq:extra_cum})
\begin{subequations}\label{eq:extra_cum_anelastic}
    \begin{align}
     \mathbf{C}_{+}^{{\Phi}} (\mathbf{r_1}, \mathbf{r_2}, t)  
     	=  \frac{1}{2}\overline{\left[\rho_0(\mathbf{r_1}) + \rho_0(\mathbf{r_2})\right]
    \hat{\Phi}(\mathbf{r_1}, t) \hat{\mathbf{\Psi}}(\mathbf{r_2},t)}, \\ 
     \mathbf{C}_{+}^u(\mathbf{r_1}, \mathbf{r_2}, t)  
     	=  \frac{1}{2}\overline{\left[\rho_0(\mathbf{r_1}) + \rho_0(\mathbf{r_2})\right]
     \hat{\mathbf{\Psi}}(\mathbf{r_2}, t) \otimes \hat{\mathbf{u}} (\mathbf{r_1},t)}, 
    \end{align}
\end{subequations}
and introduce the corresponding quantities, $\mathbf{C}_{-}$,  $\mathbf{C}_{+}^{{\Phi}}$, and $\mathbf{C}_{-}^u$ with the density difference instead of the sum, as in 
\begin{equation}
    \mathbf{C}_{-}(\mathbf{r_1}, \mathbf{r_2}, t) = 
    	\frac{1}{2}\overline{\left[\rho_0(\mathbf{r_1}) - \rho_0(\mathbf{r_2})\right]
    \hat{\mathbf{\Psi}}(\mathbf{r_1}, t) \otimes \hat{\mathbf{\Psi}}(\mathbf{r_2}, t)}.
\end{equation}
Some algebra then gives the equations of motion for the second cumulant:
\begin{subequations}\label{eq:cum2_anelastic_tensor}
    \begin{multline}
        \frac{\partial}{\partial t}\mathbf{C}_{\pm}(\mathbf{r_1},\mathbf{r_2},t) 
        + \nabla_{\mathbf{r_1}} \cdot [\mathbf{C}_{\pm}(\mathbf{r_1},\mathbf{r_2},t) \otimes \overline{\mathbf{u}}^*(\mathbf{r_1},t) ] 
        + \nabla_{\mathbf{r_2}} \cdot [\mathbf{C}_{\pm}^T(\mathbf{r_1},\mathbf{r_2},t) \otimes \overline{\mathbf{u}}^*(\mathbf{r_2},t)] = \\
        -\mathbf{C}_{\pm}^u(\mathbf{r_1},\mathbf{r_2},t) \left[\nabla \overline{\Psi}^*(\mathbf{r_2},t)\right]^T  
        -\nabla \overline{\Psi}^*(\mathbf{r_2},t) \mathbf{C}_{\pm}^u(\mathbf{r_1},\mathbf{r_2},t)  \\
        -\frac{1}{2} \mathbf{C}_{\pm}(\mathbf{r_1},\mathbf{r_2},t) \left[ \nabla \cdot \overline{\mathbf{u}}^*(\mathbf{r_1,t}) + \nabla \cdot \overline{\mathbf{u}}^*(\mathbf{r_2,t}) \right] \\
        -\frac{1}{2} \mathbf{C_{\mp}}(\mathbf{r_1},\mathbf{r_2},t) \left[ \nabla \cdot \overline{\mathbf{u}}^* (\mathbf{r_1,t}) - \nabla \cdot \overline{\mathbf{u}}^* (\mathbf{r_2,t}) \right]  \\
       + \mathbf{L}\, \mathbf{C}_{\pm}(\mathbf{r_1},\mathbf{r_2},t)  + \mathbf{C}_{\pm}(\mathbf{r_1},\mathbf{r_2},t)\,\mathbf{L}^T 
       + \nabla_{\mathbf{r_1}} \mathbf{C}_\pm^{\Phi} (\mathbf{r_1},\mathbf{r_2},t) + \nabla_{\mathbf{r_2}} \mathbf{C}_\pm^{\Phi} (\mathbf{r_2},\mathbf{r_1},t)\\
     + \mbox{third-order term} 
     \end{multline}
     \begin{align}
        \nabla_{\mathbf{r_2}} \cdot \left[ \mathbf{C}_{+}^u (\mathbf{r_1},\mathbf{r_2},t) + \mathbf{C}_{-}^u(\mathbf{r_1},\mathbf{r_2},t) \right] & = 0, \notag \\
        \nabla_{\mathbf{r_1}} \cdot \left [\mathbf{C}_{+}^u(\mathbf{r_1},\mathbf{r_2},t) - \mathbf{C}_{-}^u(\mathbf{r_1},\mathbf{r_2},t) \right] & = 0.
    \end{align}
\end{subequations}

Truncation of the cumulant expansion at second order (CE2) consists of equations (\ref{eq:anelastic_cum1_bis}) and (\ref{eq:cum2_anelastic_tensor}), neglecting the third-order term. \citet{clark1995} have derived the evolution equation of the single-time two-point covariance tensor for the compressible Navier-Stokes equation, yielding a more general version of (\ref{eq:cum2_anelastic_tensor}). 

\section{Eddy absorption in the QL approximation}\label{a:QL-large}

\begin{figure}[hb]
  \noindent\includegraphics[width=\textwidth]{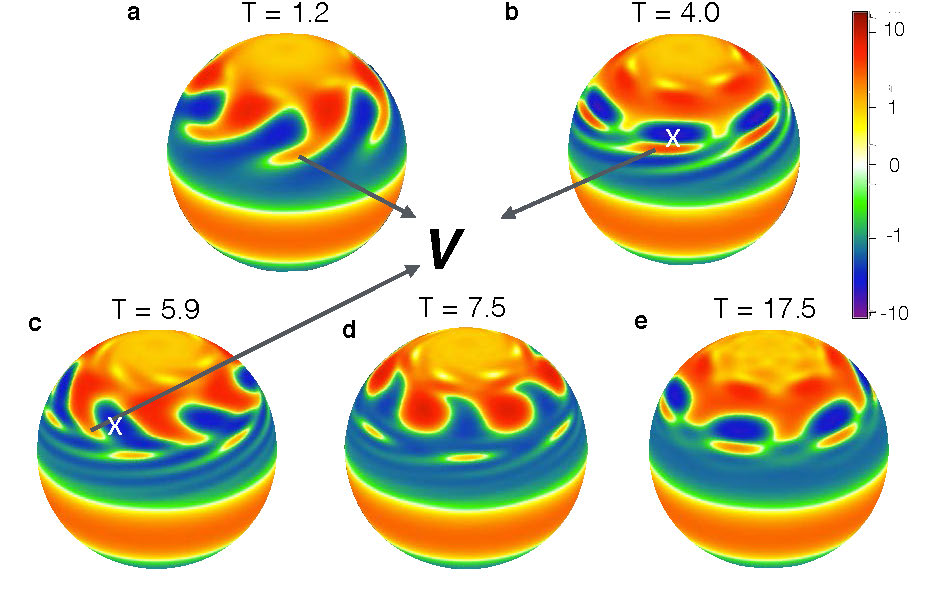}\\
  \caption{Evolution of relative vorticity in a QL simulation of the larger-amplitude eddies ($\epsilon = 6$) and an Earth-like setting ($\mathrm{Ro} = 0.06$), for which the fully nonlinear simulation is shown in Fig.~\ref{fig:EarthLike}. White X's mark the centres of what would become cats' eyes in the fully nonlinear simulation. The locations labeled by V follow a positive vorticity anomaly that detaches from an initial wave lobe.}\label{fig:EarthLikeQL}
\end{figure}

To illustrate more clearly why CE2 fails to capture the absorption of larger-amplitude eddies, we perform QL simulations of the barotropic vorticity equation (\ref{eq:barotropic}), eliminating eddy--eddy interactions as in the QL approximation (\ref{eq:QL}) of the Boussinesq equations. The relative vorticity in the QL simulation is shown in Fig.~\ref{fig:EarthLikeQL}. The positive vorticity anomaly (labeled V) that detaches from a wave crest is initially sheared by the mean flow, leading to momentum fluxes that strengthen the mean flow and to a decrease of the eddy kinetic energy (Orr mechanism). It is then advected around the centre of the cats' eye (labeled X). But instead of rolling up into a small-scale filament as it does in the fully nonlinear simulation (cf.\ Fig.~\ref{fig:EarthLike}), it moves to the western side of the eye, where it joins the wave lobe with positive vorticity west of the one from where it originated ($T=5.9$). An analogous description applies to the negative vorticity anomaly inside the eye. This leads at $T=5.9$ to vorticity anomalies that have a southeast--northwest tilt of phase lines, consistent with an equatorward eddy momentum flux (reflection phase) and an increase of EKE because the mean-flow shear goes against the tilt (Orr mechanism). The vorticity map after two cycles of absorption and reflection ($T=17.5$) is very similar to the initial one ($T=4$). Differences arise from wave absorption occurring through weakly nonlinear processes and, to a lesser extent, from hyperviscosity. This is in sharp contrast to the full simulation, in which filamentation takes place, eventually leading to irreversibility (cf.\ $T=4$ and $T=17.5$ of Fig.~\ref{fig:EarthLike}). 

{\clearpage}
\section*{References}
\begingroup
\renewcommand{\section}[2]{}
\bibliographystyle{ametsoc}
\bibliography{references}
\endgroup

\end{document}